\documentclass[conference,a4paper]{IEEEtran}
\usepackage{graphicx} 
\usepackage{array}
\usepackage{arydshln}
\usepackage{xcolor,colortbl}
\usepackage{cite}
\usepackage{tikz,  pgfplots}
\usepackage{mathrsfs}
\usepackage{footmisc}
\usepackage{amsmath}
\usepackage{amssymb}
\usepackage{amsfonts}
\usepackage{amsthm}
\usepackage{mathrsfs}
\usepackage{xspace}
\usepackage{bm}
\usepackage{fancyref}
\usepackage{textcomp}
\usepackage[Symbol]{upgreek}
\usepackage{mleftright}
\usepackage{enumerate}
\usepackage{mathtools}

\makeatletter
\if@twocolumn
\newcommand{\mysetminus}{\!\setminus\!} 
\else
\newcommand{\mysetminus}{\!\setminus\!} 
\fi
\makeatother

\newcommand{\safemath}[2]{\newcommand{#1}{\ensuremath{#2}\xspace}}

\newcommand{\ssa}{\mathsf{a}}
\newcommand{\ssb}{\mathsf{b}}
\newcommand{\ssc}{\mathsf{c}}
\newcommand{\ssd}{\mathsf{d}}
\newcommand{\sse}{\mathsf{e}}
\newcommand{\ssf}{\mathsf{f}}
\newcommand{\ssg}{\mathsf{g}}
\newcommand{\ssh}{\mathsf{h}}
\newcommand{\ssi}{\mathsf{i}}
\newcommand{\ssj}{\mathsf{j}}
\newcommand{\ssk}{\mathsf{k}}
\newcommand{\ssl}{\mathsf{l}}
\newcommand{\ssm}{\mathsf{m}}
\newcommand{\ssn}{\mathsf{n}}
\newcommand{\sso}{\mathsf{o}}
\newcommand{\ssp}{\mathsf{p}}
\newcommand{\ssq}{\mathsf{q}}
\newcommand{\ssr}{\mathsf{r}}
\newcommand{\sss}{\mathsf{s}}
\newcommand{\sst}{\mathsf{t}}
\newcommand{\ssu}{\mathsf{u}}
\newcommand{\ssv}{\mathsf{v}}
\newcommand{\ssw}{\mathsf{w}}
\newcommand{\ssx}{\mathsf{x}}
\newcommand{\ssy}{\mathsf{y}}
\newcommand{\ssz}{\mathsf{z}}

\safemath{\bmsa}{\bm{\ssa}}
\safemath{\bmsb}{\bm{\ssb}}
\safemath{\bmsc}{\bm{\ssc}}
\safemath{\bmsd}{\bm{\ssd}}
\safemath{\bmse}{\bm{\sse}}
\safemath{\bmsf}{\bm{\ssf}}
\safemath{\bmsg}{\bm{\ssg}}
\safemath{\bmsh}{\bm{\ssh}}
\safemath{\bmsi}{\bm{\ssi}}
\safemath{\bmsj}{\bm{\ssj}}
\safemath{\bmsk}{\bm{\ssk}}
\safemath{\bmsl}{\bm{\ssl}}
\safemath{\bmsm}{\bm{\ssm}}
\safemath{\bmsn}{\bm{\ssn}}
\safemath{\bmso}{\bm{\sso}}
\safemath{\bmsp}{\bm{\ssp}}
\safemath{\bmsq}{\bm{\ssq}}
\safemath{\bmsr}{\bm{\ssr}}
\safemath{\bmss}{\bm{\sss}}
\safemath{\bmst}{\bm{\sst}}
\safemath{\bmsu}{\bm{\ssu}}
\safemath{\bmsv}{\bm{\ssv}}
\safemath{\bmsw}{\bm{\ssw}}
\safemath{\bmsx}{\bm{\ssx}}
\safemath{\bmsy}{\bm{\ssy}}
\safemath{\bmsz}{\bm{\ssz}}

\bmdefine{\bmualphad}{\upalpha}
\bmdefine{\bmubetad}{\upbeta}
\bmdefine{\bmuchid}{\upchi}
\bmdefine{\bmudeltad}{\updelta}
\bmdefine{\bmuepsilond}{\upepsilon}
\bmdefine{\bmuvarepsilond}{\upvarepsilon}
\bmdefine{\bmuetad}{\upeta}
\bmdefine{\bmugammad}{\upgamma}
\bmdefine{\bmuiotad}{\upiota}
\bmdefine{\bmukappad}{\upkappa}
\bmdefine{\bmulambdad}{\uplambda}
\bmdefine{\bmumud}{\upmu}
\bmdefine{\bmunud}{\upnu}
\bmdefine{\bmuomegad}{\upomega}
\bmdefine{\bmuphid}{\upphi}
\bmdefine{\bmuvarphid}{\upvarphi}
\bmdefine{\bmupid}{\uppi}
\bmdefine{\bmuvarpid}{\upvarpi}
\bmdefine{\bmupsid}{\uppsi}
\bmdefine{\bmurhod}{\uprho}
\bmdefine{\bmuvarrhod}{\upvarrho}
\bmdefine{\bmusigmad}{\upsigma}
\bmdefine{\bmuvarsigmad}{\upvarsigma}
\bmdefine{\bmutaud}{\uptau}
\bmdefine{\bmuthetad}{\uptheta}
\bmdefine{\bmuvarthetad}{\upvartheta}
\bmdefine{\bmuupsilond}{\upupsilon}
\bmdefine{\bmuxid}{\upxi}
\bmdefine{\bmuzetad}{\upzeta}

\safemath{\bmua}{\mathbf{a}}
\safemath{\bmub}{\mathbf{b}}
\safemath{\bmuc}{\mathbf{c}}
\safemath{\bmud}{\mathbf{d}}
\safemath{\bmue}{\mathbf{e}}
\safemath{\bmuf}{\mathbf{f}}
\safemath{\bmug}{\mathbf{g}}
\safemath{\bmuh}{\mathbf{h}}
\safemath{\bmui}{\mathbf{i}}
\safemath{\bmuj}{\mathbf{j}}
\safemath{\bmuk}{\mathbf{k}}
\safemath{\bmul}{\mathbf{l}}
\safemath{\bmum}{\mathbf{m}}
\safemath{\bmun}{\mathbf{n}}
\safemath{\bmuo}{\mathbf{o}}
\safemath{\bmup}{\mathbf{p}}
\safemath{\bmuq}{\mathbf{q}}
\safemath{\bmur}{\mathbf{r}}
\safemath{\bmus}{\mathbf{s}}
\safemath{\bmut}{\mathbf{t}}
\safemath{\bmuu}{\mathbf{u}}
\safemath{\bmuv}{\mathbf{v}}
\safemath{\bmuw}{\mathbf{w}}
\safemath{\bmux}{\mathbf{x}}
\safemath{\bmuy}{\mathbf{y}}
\safemath{\bmuz}{\mathbf{z}}

\safemath{\bmualpha}{\bmualphad}
\safemath{\bmubeta}{\bmubetad}
\safemath{\bmuchi}{\bumchid}
\safemath{\bmudelta}{\bmudeltad}
\safemath{\bmuepsilon}{\bmuepsilond}
\safemath{\bmuvarepsilon}{\bmuvarepsilond}
\safemath{\bmueta}{\bmuetad}
\safemath{\bmugamma}{\bmugammad}
\safemath{\bmuiota}{\bmuiotad}
\safemath{\bmukappa}{\bmukappad}
\safemath{\bmulambda}{\bmulambdad}
\safemath{\bmumu}{\bmumud}
\safemath{\bmunu}{\bmunud}
\safemath{\bmuomega}{\bmuomegad}
\safemath{\bmuphi}{\bmuphid}
\safemath{\bmuvarphi}{\bmuvarphid}
\safemath{\bmupi}{\bmupid}
\safemath{\bmuvarpi}{\bmuvarpid}
\safemath{\bmupsi}{\bmupsid}
\safemath{\bmurho}{\bmurhod}
\safemath{\bmuvarrho}{\bmuvarrhod}
\safemath{\bmusigma}{\bmusigmad}
\safemath{\bmuvarsigma}{\bmuvarsigmad}
\safemath{\bmutau}{\bmutaud}
\safemath{\bmutheta}{\bmuthetad}
\safemath{\bmuvartheta}{\bmuvarthetad}
\safemath{\bmuupsilon}{\bmuupsilond}
\safemath{\bmuxi}{\bmuxid}
\safemath{\bmuzeta}{\bmuzetad}

\bmdefine{\bmiad}{a}
\bmdefine{\bmibd}{b}
\bmdefine{\bmicd}{c}
\bmdefine{\bmidd}{d}
\bmdefine{\bmied}{e}
\bmdefine{\bmifd}{f}
\bmdefine{\bmigd}{g}
\bmdefine{\bmihd}{h}
\bmdefine{\bmiid}{i}
\bmdefine{\bmijd}{j}
\bmdefine{\bmikd}{k}
\bmdefine{\bmild}{l}
\bmdefine{\bmimd}{m}
\bmdefine{\bmind}{n}
\bmdefine{\bmiod}{o}
\bmdefine{\bmipd}{p}
\bmdefine{\bmiqd}{q}
\bmdefine{\bmird}{r}
\bmdefine{\bmisd}{s}
\bmdefine{\bmitd}{t}
\bmdefine{\bmiud}{u}
\bmdefine{\bmivd}{v}
\bmdefine{\bmiwd}{w}
\bmdefine{\bmixd}{x}
\bmdefine{\bmiyd}{y}
\bmdefine{\bmizd}{z}

\bmdefine{\bmialphad}{\alpha}
\bmdefine{\bmibetad}{\beta}
\bmdefine{\bmichid}{\chi}
\bmdefine{\bmideltad}{\delta}
\bmdefine{\bmiepsilond}{\epsilon}
\bmdefine{\bmivarepsilond}{\varepsilon}
\bmdefine{\bmietad}{\eta}
\bmdefine{\bmigammad}{\gamma}
\bmdefine{\bmiiotad}{\iota}
\bmdefine{\bmikappad}{\kappa}
\bmdefine{\bmivarkappad}{\varkappa}
\bmdefine{\bmilambdad}{\lambda}
\bmdefine{\bmimud}{\mu}
\bmdefine{\bminud}{\nu}
\bmdefine{\bmiomegad}{\omega}
\bmdefine{\bmiphid}{\phi}
\bmdefine{\bmivarphid}{\varphi}
\bmdefine{\bmipid}{\pi}
\bmdefine{\bmivarpid}{\varpi}
\bmdefine{\bmipsid}{\psi}
\bmdefine{\bmirhod}{\rho}
\bmdefine{\bmivarrhod}{\varrho}
\bmdefine{\bmisigmad}{\sigma}
\bmdefine{\bmivarsigmad}{\varsigma}
\bmdefine{\bmitaud}{\tau}
\bmdefine{\bmithetad}{\theta}
\bmdefine{\bmivarthetad}{\vartheta}
\bmdefine{\bmiupsilond}{\upsilon}
\bmdefine{\bmixid}{\xi}
\bmdefine{\bmizetad}{\zeta}

\safemath{\bmia}{\bmiad}
\safemath{\bmib}{\bmibd}
\safemath{\bmic}{\bmicd}
\safemath{\bmid}{\bmidd}
\safemath{\bmie}{\bmied}
\safemath{\bmif}{\bmifd}
\safemath{\bmig}{\bmigd}
\safemath{\bmih}{\bmihd}
\safemath{\bmii}{\bmiid}
\safemath{\bmij}{\bmijd}
\safemath{\bmik}{\bmikd}
\safemath{\bmil}{\bmild}
\safemath{\bmim}{\bmimd}
\safemath{\bmin}{\bmind}
\safemath{\bmio}{\bmiod}
\safemath{\bmip}{\bmipd}
\safemath{\bmiq}{\bmiqd}
\safemath{\bmir}{\bmird}
\safemath{\bmis}{\bmisd}
\safemath{\bmit}{\bmitd}
\safemath{\bmiu}{\bmiud}
\safemath{\bmiv}{\bmivd}
\safemath{\bmiw}{\bmiwd}
\safemath{\bmix}{\bmixd}
\safemath{\bmiy}{\bmiyd}
\safemath{\bmiz}{\bmizd}

\safemath{\bmialpha}{\bmialphad}
\safemath{\bmibeta}{\bmibetad}
\safemath{\bmichi}{\bmichid}
\safemath{\bmidelta}{\bmideltad}
\safemath{\bmiepsilon}{\bmiepsilond}
\safemath{\bmivarepsilon}{\bmivarepsilond}
\safemath{\bmieta}{\bmietad}
\safemath{\bmigamma}{\bmigammad}
\safemath{\bmiiota}{\bmiiotad}
\safemath{\bmikappa}{\bmikappad}
\safemath{\bmivarkappa}{\bmivarkappad}
\safemath{\bmilambda}{\bmilambdad}
\safemath{\bmimu}{\bmimud}
\safemath{\bminu}{\bminud}
\safemath{\bmiomega}{\bmiomegad}
\safemath{\bmiphi}{\bmiphid}
\safemath{\bmivarphi}{\bmivarphid}
\safemath{\bmipi}{\bmipid}
\safemath{\bmivarpi}{\bmivarpid}
\safemath{\bmipsi}{\bmipsid}
\safemath{\bmirho}{\bmirhod}
\safemath{\bmivarrho}{\bmivarrhod}
\safemath{\bmisigma}{\bmisigmad}
\safemath{\bmivarsigma}{\bmivarsigmad}
\safemath{\bmitau}{\bmitaud}
\safemath{\bmitheta}{\bmithetad}
\safemath{\bmivartheta}{\bmivarthetad}
\safemath{\bmiupsilon}{\bmiupsilond}
\safemath{\bmixi}{\bmixid}
\safemath{\bmizeta}{\bmizetad}

\bmdefine{\bmuDeltad}{\Updelta}
\bmdefine{\bmuGammad}{\Upgamma}
\bmdefine{\bmuLambdad}{\Uplambda}
\bmdefine{\bmuOmegad}{\Upomega}
\bmdefine{\bmuPhid}{\Upphi}
\bmdefine{\bmuPid}{\Uppi}
\bmdefine{\bmuPsid}{\Uppsi}
\bmdefine{\bmuSigmad}{\Upsigma}
\bmdefine{\bmuThetad}{\Uptheta}
\bmdefine{\bmuUpsilond}{\Upupsilon}
\bmdefine{\bmuXid}{\Upxi}

\safemath{\bmuA}{\mathbf{A}}
\safemath{\bmuB}{\mathbf{B}}
\safemath{\bmuC}{\mathbf{C}}
\safemath{\bmuD}{\mathbf{D}}
\safemath{\bmuE}{\mathbf{E}}
\safemath{\bmuF}{\mathbf{F}}
\safemath{\bmuG}{\mathbf{G}}
\safemath{\bmuH}{\mathbf{H}}
\safemath{\bmuI}{\mathbf{I}}
\safemath{\bmuJ}{\mathbf{J}}
\safemath{\bmuK}{\mathbf{K}}
\safemath{\bmuL}{\mathbf{L}}
\safemath{\bmuM}{\mathbf{M}}
\safemath{\bmuN}{\mathbf{N}}
\safemath{\bmuO}{\mathbf{O}}
\safemath{\bmuP}{\mathbf{P}}
\safemath{\bmuQ}{\mathbf{Q}}
\safemath{\bmuR}{\mathbf{R}}
\safemath{\bmuS}{\mathbf{S}}
\safemath{\bmuT}{\mathbf{T}}
\safemath{\bmuU}{\mathbf{U}}
\safemath{\bmuV}{\mathbf{V}}
\safemath{\bmuW}{\mathbf{W}}
\safemath{\bmuX}{\mathbf{X}}
\safemath{\bmuY}{\mathbf{Y}}
\safemath{\bmuZ}{\mathbf{Z}}

\safemath{\bmuZero}{\mathbf{0}}
\safemath{\bmuOne}{\mathbf{1}}

\safemath{\bmuDelta}{\bmuDeltad}
\safemath{\bmuGamma}{\bmuGammad}
\safemath{\bmuLambda}{\bmuLambdad}
\safemath{\bmuOmega}{\bmuOmegad}
\safemath{\bmuPhi}{\bmuPhid}
\safemath{\bmuPi}{\bmuPid}
\safemath{\bmuPsi}{\bmuPsid}
\safemath{\bmuSigma}{\bmuSigmad}
\safemath{\bmuTheta}{\bmuThetad}
\safemath{\bmuUpsilon}{\bmuUpsilond}
\safemath{\bmuXi}{\bmuXid}

\bmdefine{\bmiAd}{A}
\bmdefine{\bmiBd}{B}
\bmdefine{\bmiCd}{C}
\bmdefine{\bmiDd}{D}
\bmdefine{\bmiEd}{E}
\bmdefine{\bmiFd}{F}
\bmdefine{\bmiGd}{G}
\bmdefine{\bmiHd}{H}
\bmdefine{\bmiId}{I}
\bmdefine{\bmiJd}{J}
\bmdefine{\bmiKd}{K}
\bmdefine{\bmiLd}{L}
\bmdefine{\bmiMd}{M}
\bmdefine{\bmiOd}{N}
\bmdefine{\bmiPd}{O}
\bmdefine{\bmiQd}{P}
\bmdefine{\bmiRd}{R}
\bmdefine{\bmiSd}{S}
\bmdefine{\bmiTd}{T}
\bmdefine{\bmiUd}{U}
\bmdefine{\bmiVd}{V}
\bmdefine{\bmiWd}{W}
\bmdefine{\bmiXd}{X}
\bmdefine{\bmiYd}{Y}
\bmdefine{\bmiZd}{Z}

\bmdefine{\bmiDeltad}{\Delta}
\bmdefine{\bmiGammad}{\Gamma}
\bmdefine{\bmiLambdad}{\Lambda}
\bmdefine{\bmiOmegad}{\Omega}
\bmdefine{\bmiPhid}{\Phi}
\bmdefine{\bmiPid}{\Pi}
\bmdefine{\bmiPsid}{\Psi}
\bmdefine{\bmiSigmad}{\Sigma}
\bmdefine{\bmiThetad}{\Theta}
\bmdefine{\bmiUpsilond}{\Upsilon}
\bmdefine{\bmiXid}{\Xi}

\safemath{\bmiA}{\bmiAd}
\safemath{\bmiB}{\bmiBd}
\safemath{\bmiC}{\bmiCd}
\safemath{\bmiD}{\bmiDd}
\safemath{\bmiE}{\bmiEd}
\safemath{\bmiF}{\bmiFd}
\safemath{\bmiG}{\bmiGd}
\safemath{\bmiH}{\bmiHd}
\safemath{\bmiI}{\bmiId}
\safemath{\bmiJ}{\bmiJd}
\safemath{\bmiK}{\bmiKd}
\safemath{\bmiL}{\bmiLd}
\safemath{\bmiM}{\bmiMd}
\safemath{\bmiN}{\bmiNd}
\safemath{\bmiO}{\bmiOd}
\safemath{\bmiP}{\bmiPd}
\safemath{\bmiQ}{\bmiQd}
\safemath{\bmiR}{\bmiRd}
\safemath{\bmiS}{\bmiSd}
\safemath{\bmiT}{\bmiTd}
\safemath{\bmiU}{\bmiUd}
\safemath{\bmiV}{\bmiVd}
\safemath{\bmiW}{\bmiWd}
\safemath{\bmiX}{\bmiXd}
\safemath{\bmiY}{\bmiYd}
\safemath{\bmiZ}{\bmiZd}

\safemath{\bmiDelta}{\bmiDeltad}
\safemath{\bmiGamma}{\bmiGammad}
\safemath{\bmiLambda}{\bmiLambdad}
\safemath{\bmiOmega}{\bmiOmegad}
\safemath{\bmiPhi}{\bmiPhid}
\safemath{\bmiPi}{\bmiPid}
\safemath{\bmiPsi}{\bmiPsid}
\safemath{\bmiSigma}{\bmiSigmad}
\safemath{\bmiTheta}{\bmiThetad}
\safemath{\bmiUpsilon}{\bmiUpsilond}
\safemath{\bmiXi}{\bmiXid}

\safemath{\evA}{\mathcal{A}}
\safemath{\evB}{\mathcal{B}}
\safemath{\evC}{\mathcal{C}}
\safemath{\evD}{\mathcal{D}}
\safemath{\evE}{\mathcal{E}}
\safemath{\evF}{\mathcal{F}}
\safemath{\evG}{\mathcal{G}}
\safemath{\evH}{\mathcal{H}}
\safemath{\evI}{\mathcal{I}}
\safemath{\evJ}{\mathcal{J}}
\safemath{\evK}{\mathcal{K}}
\safemath{\evL}{\mathcal{L}}
\safemath{\evM}{\mathcal{M}}
\safemath{\evN}{\mathcal{N}}
\safemath{\evO}{\mathcal{O}}
\safemath{\evP}{\mathcal{P}}
\safemath{\evQ}{\mathcal{Q}}
\safemath{\evR}{\mathcal{R}}
\safemath{\evS}{\mathcal{S}}
\safemath{\evT}{\mathcal{T}}
\safemath{\evU}{\mathcal{U}}
\safemath{\evV}{\mathcal{V}}
\safemath{\evW}{\mathcal{W}}
\safemath{\evX}{\mathcal{X}}
\safemath{\evY}{\mathcal{Y}}
\safemath{\evZ}{\mathcal{Z}}

\safemath{\setA}{\mathcal{A}}
\safemath{\setB}{\mathcal{B}}
\safemath{\setC}{\mathcal{C}}
\safemath{\setD}{\mathcal{D}}
\safemath{\setE}{\mathcal{E}}
\safemath{\setF}{\mathcal{F}}
\safemath{\setG}{\mathcal{G}}
\safemath{\setH}{\mathcal{H}}
\safemath{\setI}{\mathcal{I}}
\safemath{\setJ}{\mathcal{J}}
\safemath{\setK}{\mathcal{K}}
\safemath{\setL}{\mathcal{L}}
\safemath{\setM}{\mathcal{M}}
\safemath{\setN}{\mathcal{N}}
\safemath{\setO}{\mathcal{O}}
\safemath{\setP}{\mathcal{P}}
\safemath{\setQ}{\mathcal{Q}}
\safemath{\setR}{\mathcal{R}}
\safemath{\setS}{\mathcal{S}}
\safemath{\setT}{\mathcal{T}}
\safemath{\setU}{\mathcal{U}}
\safemath{\setV}{\mathcal{V}}
\safemath{\setW}{\mathcal{W}}
\safemath{\setX}{\mathcal{X}}
\safemath{\setY}{\mathcal{Y}}
\safemath{\setZ}{\mathcal{Z}}
\safemath{\emptySet}{\varnothing}

\safemath{\colA}{\mathscr{A}}
\safemath{\colB}{\mathscr{B}}
\safemath{\colC}{\mathscr{C}}
\safemath{\colD}{\mathscr{D}}
\safemath{\colE}{\mathscr{E}}
\safemath{\colF}{\mathscr{F}}
\safemath{\colG}{\mathscr{G}}
\safemath{\colH}{\mathscr{H}}
\safemath{\colI}{\mathscr{I}}
\safemath{\colJ}{\mathscr{J}}
\safemath{\colK}{\mathscr{K}}
\safemath{\colL}{\mathscr{L}}
\safemath{\colM}{\mathscr{M}}
\safemath{\colN}{\mathscr{N}}
\safemath{\colO}{\mathscr{O}}
\safemath{\colP}{\mathscr{P}}
\safemath{\colQ}{\mathscr{Q}}
\safemath{\colR}{\mathscr{R}}
\safemath{\colS}{\mathscr{S}}
\safemath{\colT}{\mathscr{T}}
\safemath{\colU}{\mathscr{U}}
\safemath{\colV}{\mathscr{V}}
\safemath{\colW}{\mathscr{W}}
\safemath{\colX}{\mathscr{X}}
\safemath{\colY}{\mathscr{Y}}
\safemath{\colZ}{\mathscr{Z}}

\safemath{\opA}{\operatorname{A}}
\safemath{\opB}{\operatorname{B}}
\safemath{\opC}{\operatorname{C}}
\safemath{\opD}{\operatorname{D}}
\safemath{\opE}{\operatorname{E}}
\safemath{\opF}{\operatorname{F}}
\safemath{\opG}{\operatorname{G}}
\safemath{\opH}{\operatorname{H}}
\safemath{\opI}{\operatorname{I}}
\safemath{\opJ}{\operatorname{J}}
\safemath{\opK}{\operatorname{K}}
\safemath{\opL}{\operatorname{L}}
\safemath{\opM}{\operatorname{M}}
\safemath{\opN}{\operatorname{N}}
\safemath{\opO}{\operatorname{O}}
\safemath{\opP}{\operatorname{P}}
\safemath{\opQ}{\operatorname{Q}}
\safemath{\opR}{\operatorname{R}}
\safemath{\opS}{\operatorname{S}}
\safemath{\opT}{\operatorname{T}}
\safemath{\opU}{\operatorname{U}}
\safemath{\opV}{\operatorname{V}}
\safemath{\opW}{\operatorname{W}}
\safemath{\opX}{\operatorname{X}}
\safemath{\opY}{\operatorname{Y}}
\safemath{\opZ}{\operatorname{Z}}
\safemath{\opZero}{\operatorname{O}}
\safemath{\identityop}{\opI}

\safemath{\sca}{a}
\safemath{\scb}{b}
\safemath{\scc}{c}
\safemath{\scd}{d}
\safemath{\sce}{e}
\safemath{\scf}{f}
\safemath{\scg}{g}
\safemath{\sch}{h}
\safemath{\sci}{i}
\safemath{\scj}{j}
\safemath{\sck}{k}
\safemath{\scl}{l}
\safemath{\scm}{m}
\safemath{\scn}{n}
\safemath{\sco}{o}
\safemath{\scp}{p}
\safemath{\scq}{q}
\safemath{\scr}{r}
\safemath{\scs}{s}
\safemath{\sct}{t}
\safemath{\scu}{u}
\safemath{\scv}{v}
\safemath{\scw}{w}
\safemath{\scx}{x}
\safemath{\scy}{y}
\safemath{\scz}{z}

\safemath{\scA}{A}
\safemath{\scB}{B}
\safemath{\scC}{C}
\safemath{\scD}{D}
\safemath{\scE}{E}
\safemath{\scF}{F}
\safemath{\scG}{G}
\safemath{\scH}{H}
\safemath{\scI}{I}
\safemath{\scJ}{J}
\safemath{\scK}{K}
\safemath{\scL}{L}
\safemath{\scM}{M}
\safemath{\scN}{N}
\safemath{\scO}{O}
\safemath{\scP}{P}
\safemath{\scQ}{Q}
\safemath{\scR}{R}
\safemath{\scS}{S}
\safemath{\scT}{T}
\safemath{\scU}{U}
\safemath{\scV}{V}
\safemath{\scW}{W}
\safemath{\scX}{X}
\safemath{\scY}{Y}
\safemath{\scZ}{Z}

\safemath{\scalpha}{\alpha}
\safemath{\scbeta}{\beta}
\safemath{\scchi}{\chi}
\safemath{\scdelta}{\delta}
\safemath{\scepsilon}{\epsilon}
\safemath{\scvarepsilon}{\varepsilon}
\safemath{\sceta}{\eta}
\safemath{\scgamma}{\gamma}
\safemath{\sciota}{\iota}
\safemath{\sckappa}{\kappa}
\safemath{\scvarkappa}{\varkappa}
\safemath{\sclambda}{\lambda}
\safemath{\scmu}{\mu}
\safemath{\scnu}{\nu}
\safemath{\scomega}{\omega}
\safemath{\scphi}{\phi}
\safemath{\scvarphi}{\varphi}
\safemath{\scpi}{\pi}
\safemath{\scvarpi}{\varpi}
\safemath{\scpsi}{\psi}
\safemath{\scrho}{\rho}
\safemath{\scvarrho}{\varrho}
\safemath{\scsigma}{\sigma}
\safemath{\scvarsigma}{\varsigma}
\safemath{\sctau}{\tau}
\safemath{\sctheta}{\theta}
\safemath{\scvartheta}{\vartheta}
\safemath{\scupsilon}{\upsilon}
\safemath{\scxi}{\xi}
\safemath{\sczeta}{\zeta}

\safemath{\veca}{{\boldsymbol{a}}}
\safemath{\vecb}{{\boldsymbol{b}}}
\safemath{\vecc}{{\boldsymbol{c}}}
\safemath{\vecd}{{\boldsymbol{d}}}
\safemath{\vece}{{\boldsymbol{e}}}
\safemath{\vecf}{{\boldsymbol{f}}}
\safemath{\vecg}{{\boldsymbol{g}}}
\safemath{\vech}{{\boldsymbol{h}}}
\safemath{\veci}{{\boldsymbol{i}}}
\safemath{\vecj}{{\boldsymbol{j}}}
\safemath{\veck}{{\boldsymbol{k}}}
\safemath{\vecl}{{\boldsymbol{l}}}
\safemath{\vecm}{{\boldsymbol{m}}}
\safemath{\vecn}{{\boldsymbol{n}}}
\safemath{\veco}{{\boldsymbol{o}}}
\safemath{\vecp}{{\boldsymbol{p}}}
\safemath{\vecq}{{\boldsymbol{q}}}
\safemath{\vecr}{{\boldsymbol{r}}}
\safemath{\vecs}{{\boldsymbol{s}}}
\safemath{\vect}{{\boldsymbol{t}}}
\safemath{\vecu}{{\boldsymbol{u}}}
\safemath{\vecv}{{\boldsymbol{v}}}
\safemath{\vecw}{{\boldsymbol{w}}}
\safemath{\vecx}{{\boldsymbol{x}}}
\safemath{\vecy}{{\boldsymbol{y}}}
\safemath{\vecz}{{\boldsymbol{z}}}
\safemath{\veczero}{{\boldsymbol{0}}}
\safemath{\vecone}{{\boldsymbol{1}}}

\safemath{\vecalpha}{\upalpha}
\safemath{\vecbeta}{\upbeta}
\safemath{\vecchi}{\upchi}
\safemath{\vecdelta}{\updelta}
\safemath{\vecepsilon}{\upepsilon}
\safemath{\vecvarepsilon}{\upvarepsilon}
\safemath{\veceta}{\upeta}
\safemath{\vecgamma}{\upgamma}
\safemath{\veciota}{\upiota}
\safemath{\veckappa}{\upkappa}
\safemath{\veclambda}{\uplambda}
\safemath{\vecmu}{\text{\textmu}}
\safemath{\vecnu}{\upnu}
\safemath{\vecomega}{\upomega}
\safemath{\vecphi}{\upphi}
\safemath{\vecvarphi}{\upvarphi}
\safemath{\vecpi}{\uppi}
\safemath{\vecvarpi}{\upvarpi}
\safemath{\vecpsi}{\uppsi}
\safemath{\vecrho}{\uprho}
\safemath{\vecvarrho}{\upvarrho}
\safemath{\vecsigma}{\upsigma}
\safemath{\vecvarsigma}{\upvarsigma}
\safemath{\vectau}{\uptau}
\safemath{\vectheta}{\uptheta}
\safemath{\vecvartheta}{\upvartheta}
\safemath{\vecupsilon}{\upupsilon}
\safemath{\vecxi}{\upxi}
\safemath{\veczeta}{\upzeta}

\safemath{\vecac}{a}
\safemath{\vecbc}{b}
\safemath{\veccc}{c}
\safemath{\vecdc}{d}
\safemath{\vecec}{e}
\safemath{\vecfc}{f}
\safemath{\vecgc}{g}
\safemath{\vechc}{h}
\safemath{\vecic}{i}
\safemath{\vecjc}{j}
\safemath{\veckc}{k}
\safemath{\veclc}{l}
\safemath{\vecmc}{m}
\safemath{\vecnc}{n}
\safemath{\vecoc}{o}
\safemath{\vecpc}{p}
\safemath{\vecqc}{q}
\safemath{\vecrc}{r}
\safemath{\vecsc}{s}
\safemath{\vectc}{t}
\safemath{\vecuc}{u}
\safemath{\vecvc}{v}
\safemath{\vecwc}{w}
\safemath{\vecxc}{x}
\safemath{\vecyc}{y}
\safemath{\veczc}{z}

\safemath{\matA}{{\boldsymbol{A}}}
\safemath{\matB}{{\boldsymbol{B}}}
\safemath{\matC}{{\boldsymbol{C}}}
\safemath{\matD}{{\boldsymbol{D}}}
\safemath{\matE}{{\boldsymbol{E}}}
\safemath{\matF}{{\boldsymbol{F}}}
\safemath{\matG}{{\boldsymbol{G}}}
\safemath{\matH}{{\boldsymbol{H}}}
\safemath{\matI}{{\boldsymbol{I}}}
\safemath{\matJ}{{\boldsymbol{J}}}
\safemath{\matK}{{\boldsymbol{K}}}
\safemath{\matL}{{\boldsymbol{L}}}
\safemath{\matM}{{\boldsymbol{M}}}
\safemath{\matN}{{\boldsymbol{N}}}
\safemath{\matO}{{\boldsymbol{O}}}
\safemath{\matP}{{\boldsymbol{P}}}
\safemath{\matQ}{{\boldsymbol{Q}}}
\safemath{\matR}{{\boldsymbol{R}}}
\safemath{\matS}{{\boldsymbol{S}}}
\safemath{\matT}{{\boldsymbol{T}}}
\safemath{\matU}{{\boldsymbol{U}}}
\safemath{\matV}{{\boldsymbol{V}}}
\safemath{\matW}{{\boldsymbol{W}}}
\safemath{\matX}{{\boldsymbol{X}}}
\safemath{\matY}{{\boldsymbol{Y}}}
\safemath{\matZ}{{\boldsymbol{Z}}}
\safemath{\matzero}{{\boldsymbol{0}}}

\safemath{\matDelta}{\Updelta}
\safemath{\matGamma}{\Upgammma}
\safemath{\matLambda}{\Uplambda}
\safemath{\matOmega}{\Upomega}
\safemath{\matPhi}{\Upphi}
\safemath{\matPi}{\Uppi}
\safemath{\matPsi}{\Uppsi}
\safemath{\matSigma}{\Upsigma}
\safemath{\matTheta}{\Uptheta}
\safemath{\matUpsilon}{\Upupsilon}
\safemath{\matXi}{\Upxi}

\safemath{\matidentity}{\matI}
\safemath{\vecunit}{\vece} 
\safemath{\matone}{\matO}

\safemath{\matAc}{a}
\safemath{\matBc}{b}
\safemath{\matCc}{c}
\safemath{\matDc}{d}
\safemath{\matEc}{e}
\safemath{\matFc}{f}
\safemath{\matGc}{g}
\safemath{\matHc}{h}
\safemath{\matIc}{i}
\safemath{\matJc}{j}
\safemath{\matKc}{k}
\safemath{\matLc}{l}
\safemath{\matMc}{m}
\safemath{\matNc}{n}
\safemath{\matOc}{o}
\safemath{\matPc}{p}
\safemath{\matQc}{q}
\safemath{\matRc}{r}
\safemath{\matSc}{s}
\safemath{\matTc}{t}
\safemath{\matUc}{u}
\safemath{\matVc}{v}
\safemath{\matWc}{w}
\safemath{\matXc}{x}
\safemath{\matYc}{y}
\safemath{\matZc}{z}

\safemath{\rnda}{\mathsf{a}}
\safemath{\rndb}{\mathsf{b}}
\safemath{\rndc}{\mathsf{c}}
\safemath{\rndd}{\mathsf{d}}
\safemath{\rnde}{\mathsf{e}}
\safemath{\rndf}{\mathsf{f}}
\safemath{\rndg}{\mathsf{g}}
\safemath{\rndh}{\mathsf{h}}
\safemath{\rndi}{\mathsf{i}}
\safemath{\rndj}{\mathsf{j}}
\safemath{\rndk}{\mathsf{k}}
\safemath{\rndl}{\mathsf{l}}
\safemath{\rndm}{\mathsf{m}}
\safemath{\rndn}{\mathsf{n}}
\safemath{\rndo}{\mathsf{o}}
\safemath{\rndp}{\mathsf{p}}
\safemath{\rndq}{\mathsf{q}}
\safemath{\rndr}{\mathsf{r}}
\safemath{\rnds}{\mathsf{s}}
\safemath{\rndt}{\mathsf{t}}
\safemath{\rndu}{\mathsf{u}}
\safemath{\rndv}{\mathsf{v}}
\safemath{\rndw}{\mathsf{w}}
\safemath{\rndx}{\mathsf{x}}
\safemath{\rndy}{\mathsf{y}}
\safemath{\rndz}{\mathsf{z}}

\safemath{\rndA}{\bmiA}
\safemath{\rndB}{\bmiB}
\safemath{\rndC}{\bmiC}
\safemath{\rndD}{\bmiD}
\safemath{\rndE}{\bmiE}
\safemath{\rndF}{\bmiF}
\safemath{\rndG}{\bmiG}
\safemath{\rndH}{\bmiH}
\safemath{\rndI}{\bmiI}
\safemath{\rndJ}{\bmiJ}
\safemath{\rndK}{\bmiK}
\safemath{\rndL}{\bmiL}
\safemath{\rndM}{\bmiM}
\safemath{\rndN}{\bmiN}
\safemath{\rndO}{\bmiO}
\safemath{\rndP}{\bmiP}
\safemath{\rndQ}{\bmiQ}
\safemath{\rndR}{\bmiR}
\safemath{\rndS}{\bmiS}
\safemath{\rndT}{\bmiT}
\safemath{\rndU}{\bmiU}
\safemath{\rndV}{\bmiV}
\safemath{\rndW}{\bmiW}
\safemath{\rndX}{\bmiX}
\safemath{\rndY}{\bmiY}
\safemath{\rndZ}{\bmiZ}

\safemath{\rndalpha}{\bmialpha}
\safemath{\rndbeta}{\bmibeta}
\safemath{\rndchi}{\bmichi}
\safemath{\rnddelta}{\bmidelta}
\safemath{\rndepsilon}{\bmiepsilon}
\safemath{\rndvarepsilon}{\bmivarepsilon}
\safemath{\rndeta}{\bmieta}
\safemath{\rndgamma}{\bmigamma}
\safemath{\rndiota}{\bmiiota}
\safemath{\rndkappa}{\bmikappa}
\safemath{\rndlambda}{\bmilambda}
\safemath{\rndmu}{\bmimu}
\safemath{\rndnu}{\bminu}
\safemath{\rndomega}{\bmiomega}
\safemath{\rndphi}{\bmiphi}
\safemath{\rndvarphi}{\bmivarphi}
\safemath{\rndpi}{\bmipi}
\safemath{\rndvarpi}{\bmivarpi}
\safemath{\rndpsi}{\bmipsi}
\safemath{\rndrho}{\bmirho}
\safemath{\rndvarrho}{\bmivarrho}
\safemath{\rndsigma}{\bmisigma}
\safemath{\rndvarsigma}{\bmivarsigma}
\safemath{\rndtau}{\bmitau}
\safemath{\rndtheta}{\bmitheta}
\safemath{\rndvartheta}{\bmivartheta}
\safemath{\rndupsilon}{\bmiupsilon}
\safemath{\rndxi}{\bmixi}
\safemath{\rndzeta}{\bmizeta}

\safemath{\rveca}{{\boldsymbol{\mathsf{a}}}}
\safemath{\rvecb}{{\boldsymbol{\mathsf{b}}}}
\safemath{\rvecc}{{\boldsymbol{\mathsf{c}}}}
\safemath{\rvecd}{{\boldsymbol{\mathsf{d}}}}
\safemath{\rvece}{{\boldsymbol{\mathsf{e}}}}
\safemath{\rvecf}{{\boldsymbol{\mathsf{f}}}}
\safemath{\rvecg}{{\boldsymbol{\mathsf{g}}}}
\safemath{\rvech}{{\boldsymbol{\mathsf{h}}}}
\safemath{\rveci}{{\boldsymbol{\mathsf{i}}}}
\safemath{\rvecj}{{\boldsymbol{\mathsf{j}}}}
\safemath{\rveck}{{\boldsymbol{\mathsf{k}}}}
\safemath{\rvecl}{{\boldsymbol{\mathsf{l}}}}
\safemath{\rvecm}{{\boldsymbol{\mathsf{m}}}}
\safemath{\rvecn}{{\boldsymbol{\mathsf{n}}}}
\safemath{\rveco}{{\boldsymbol{\mathsf{o}}}}
\safemath{\rvecp}{{\boldsymbol{\mathsf{p}}}}
\safemath{\rvecq}{{\boldsymbol{\mathsf{q}}}}
\safemath{\rvecr}{{\boldsymbol{\mathsf{r}}}}
\safemath{\rvecs}{{\boldsymbol{\mathsf{s}}}}
\safemath{\rvect}{{\boldsymbol{\mathsf{t}}}}
\safemath{\rvecu}{{\boldsymbol{\mathsf{u}}}}
\safemath{\rvecv}{{\boldsymbol{\mathsf{v}}}}
\safemath{\rvecw}{{\boldsymbol{\mathsf{w}}}}
\safemath{\rvecx}{{\boldsymbol{\mathsf{x}}}}
\safemath{\rvecy}{{\boldsymbol{\mathsf{y}}}}
\safemath{\rvecz}{{\boldsymbol{\mathsf{z}}}}

\safemath{\rvecalpha}{\bmualpha}
\safemath{\rvecbeta}{\bmubeta}
\safemath{\rvecchi}{\bmuchi}
\safemath{\rvecdelta}{\bmudelta}
\safemath{\rvecepsilon}{\bmuepsilon}
\safemath{\rvecvarepsilon}{\bmuvarepsilon}
\safemath{\rveceta}{\bmueta}
\safemath{\rvecgamma}{\bmugamma}
\safemath{\rveciota}{\bmuiota}
\safemath{\rveckappa}{\bmukappa}
\safemath{\rveclambda}{\bmulambda}
\safemath{\rvecmu}{\bmumu}
\safemath{\rvecnu}{\bmunu}
\safemath{\rvecomega}{\bmuomega}
\safemath{\rvecphi}{\bmuphi}
\safemath{\rvecvarphi}{\bmuvarphi}
\safemath{\rvecpi}{\bmupi}
\safemath{\rvecvarpi}{\bmuvarpi}
\safemath{\rvecpsi}{\bmupsi}
\safemath{\rvecrho}{\bmurho}
\safemath{\rvecvarrho}{\bmuvarrho}
\safemath{\rvecsigma}{\bmusigma}
\safemath{\rvecvarsigma}{\bmuvarsigma}
\safemath{\rvectau}{\bmutau}
\safemath{\rvectheta}{\bmutheta}
\safemath{\rvecvartheta}{\bmuvartheta}
\safemath{\rvecupsilon}{\bmuupsilon}
\safemath{\rvecxi}{\bmuxi}
\safemath{\rveczeta}{\bmuzeta}

\safemath{\rvecac}{\rnda}
\safemath{\rvecbc}{\rndb}
\safemath{\rveccc}{\rndc}
\safemath{\rvecdc}{\rndd}
\safemath{\rvecec}{\rnde}
\safemath{\rvecfc}{\rndf}
\safemath{\rvecgc}{\rndg}
\safemath{\rvechc}{\rndh}
\safemath{\rvecic}{\rndi}
\safemath{\rvecjc}{\rndj}
\safemath{\rveckc}{\rndk}
\safemath{\rveclc}{\rndl}
\safemath{\rvecmc}{\rndm}
\safemath{\rvecnc}{\rndn}
\safemath{\rvecoc}{\rndo}
\safemath{\rvecpc}{\rndp}
\safemath{\rvecqc}{\rndq}
\safemath{\rvecrc}{\rndr}
\safemath{\rvecsc}{\rnds}
\safemath{\rvectc}{\rndt}
\safemath{\rvecuc}{\rndu}
\safemath{\rvecvc}{\rndv}
\safemath{\rvecwc}{\rndw}
\safemath{\rvecxc}{\rndx}
\safemath{\rvecyc}{\rndy}
\safemath{\rveczc}{\rndz}

\safemath{\rmatA}{{\boldsymbol{\mathsf{A}}}}
\safemath{\rmatB}{{\boldsymbol{\mathsf{B}}}}
\safemath{\rmatC}{{\boldsymbol{\mathsf{C}}}}
\safemath{\rmatD}{{\boldsymbol{\mathsf{D}}}}
\safemath{\rmatE}{{\boldsymbol{\mathsf{E}}}}
\safemath{\rmatF}{{\boldsymbol{\mathsf{F}}}}
\safemath{\rmatG}{{\boldsymbol{\mathsf{G}}}}
\safemath{\rmatH}{{\boldsymbol{\mathsf{H}}}}
\safemath{\rmatI}{{\boldsymbol{\mathsf{I}}}}
\safemath{\rmatJ}{{\boldsymbol{\mathsf{J}}}}
\safemath{\rmatK}{{\boldsymbol{\mathsf{K}}}}
\safemath{\rmatL}{{\boldsymbol{\mathsf{L}}}}
\safemath{\rmatM}{{\boldsymbol{\mathsf{M}}}}
\safemath{\rmatN}{{\boldsymbol{\mathsf{N}}}}
\safemath{\rmatO}{{\boldsymbol{\mathsf{O}}}}
\safemath{\rmatP}{{\boldsymbol{\mathsf{P}}}}
\safemath{\rmatQ}{{\boldsymbol{\mathsf{Q}}}}
\safemath{\rmatR}{{\boldsymbol{\mathsf{R}}}}
\safemath{\rmatS}{{\boldsymbol{\mathsf{S}}}}
\safemath{\rmatT}{{\boldsymbol{\mathsf{T}}}}
\safemath{\rmatU}{{\boldsymbol{\mathsf{U}}}}
\safemath{\rmatV}{{\boldsymbol{\mathsf{V}}}}
\safemath{\rmatW}{{\boldsymbol{\mathsf{W}}}}
\safemath{\rmatX}{{\boldsymbol{\mathsf{X}}}}
\safemath{\rmatY}{{\boldsymbol{\mathsf{Y}}}}
\safemath{\rmatZ}{{\boldsymbol{\mathsf{Z}}}}
\safemath{\rmatDelta}{\bmuDelta}
\safemath{\rmatGamma}{\bmuGamma}
\safemath{\rmatLambda}{\bmuLambda}
\safemath{\rmatOmega}{\bmuOmega}
\safemath{\rmatPhi}{\bmuPhi}
\safemath{\rmatPi}{\bmuPi}
\safemath{\rmatPsi}{\bmuPsi}
\safemath{\rmatSigma}{\bmuSigma}
\safemath{\rmatTheta}{\bmuTheta}
\safemath{\rmatUpsilon}{\bmuUpsilon}
\safemath{\rmatXi}{\bmuXi}

\safemath{\rmatAc}{\rnda}
\safemath{\rmatBc}{\rndb}
\safemath{\rmatCc}{\rndc}
\safemath{\rmatDc}{\rndd}
\safemath{\rmatEc}{\rnde}
\safemath{\rmatFc}{\rndf}
\safemath{\rmatGc}{\rndg}
\safemath{\rmatHc}{\rndh}
\safemath{\rmatIc}{\rndi}
\safemath{\rmatJc}{\rndj}
\safemath{\rmatKc}{\rndk}
\safemath{\rmatLc}{\rndl}
\safemath{\rmatMc}{\rndm}
\safemath{\rmatNc}{\rndn}
\safemath{\rmatOc}{\rndo}
\safemath{\rmatPc}{\rndp}
\safemath{\rmatQc}{\rndq}
\safemath{\rmatRc}{\rndr}
\safemath{\rmatSc}{\rnds}
\safemath{\rmatTc}{\rndt}
\safemath{\rmatUc}{\rndu}
\safemath{\rmatVc}{\rndv}
\safemath{\rmatWc}{\rndw}
\safemath{\rmatXc}{\rndx}
\safemath{\rmatYc}{\rndy}
\safemath{\rmatZc}{\rndz}

\newenvironment{textbmatrix}{	\setlength{\arraycolsep}{2.5pt}%
								\big[\begin{matrix}}{\end{matrix}\big]%
								\raisebox{0.08ex}{\vphantom{M}}}

 \def\be{\begin{equation}}
 \def\ee{\end{equation}}
 \def\btm{\begin{textbmatrix}}
 \def\etm{\end{textbmatrix}}

\def\ba#1\ea{\begin{align}#1\end{align}}

\DeclareMathOperator{\adj}{adj}				

\safemath{\fun}{\scf}						

\safemath{\vrbl}{x}						
\safemath{\altvrbl}{y}						
\safemath{\aaltvrbl}{z}						

\safemath{\vvrbl}{\vecx}						
\safemath{\altvvrbl}{\vecy}						
\safemath{\aaltvvrbl}{\vecz}						

\safemath{\altfun}{\scg}
\safemath{\aaltfun}{\sch}
\safemath{\bel}{\sce}					
\safemath{\altbel}{\sce}					
\safemath{\frel}{g}					
\safemath{\altfrel}{g}					
\safemath{\dfrel}{\tilde{g}}					
\safemath{\altdfrel}{\tilde{g}}					

\safemath{\mat}{\matA}						
\safemath{\matc}{\matAc}						
\safemath{\altmat}{\matB}						
\safemath{\altmatc}{\matBc}						

\safemath{\vectr}{\vecu}						
\safemath{\vectrc}{\vecuc}						
\safemath{\altvectr}{\vecv}						
\safemath{\aaltvectr}{\vect}						
\safemath{\altvectrc}{\vecvc}						
\safemath{\genvar}{u}						
\safemath{\altgenvar}{v}						

\safemath{\rvectr}{\rvecu}						
\safemath{\rvectrc}{\rvecuc}						
\safemath{\raltvectr}{\rvecv}						
\safemath{\raaltvectr}{\rvect}						
\safemath{\raltvectrc}{\rvecvc}						
\safemath{\rgenvar}{\rndu}						
\safemath{\raltgenvar}{\rndv}						

\newcommand{\nullspace}{\setN}	 			

\newcommand{\conj}[1]{\ensuremath{#1^{*}}} 	

\newcommand{\inv}[1]{\ensuremath{#1^{-1}}} 	

\safemath{\dirac}{\delta}					
\safemath{\diracp}{\dirac(\time)}			
\safemath{\krond}{\dirac}					
\safemath{\indfun}{I}						
\safemath{\stepfun}{u}						

\safemath{\upto}{\uparrow}
\safemath{\downto}{\downarrow}
\safemath{\iu}{\mathrm{i}}							
\safemath{\maj}{\succ}
\newcommand{\dftmat}[1]{\matF_{#1}}			
\safemath{\mdft}{\dftmat{}}					
\safemath{\runity}{\beta}					
\safemath{\eval}{\lambda}					

\safemath{\veval}{\veclambda}				
\safemath{\littleo}{\sco}					

\let\im\undefined
\safemath{\re}{\Re}				
\safemath{\im}{\Im}				

\safemath{\euclidspace}{\complexset}			
\safemath{\confunspace}{\setC}				
\newcommand{\banachseqspace}[1]{l^{#1}}		
\safemath{\hilseqspace}{\banachseqspace{2}}	
\newcommand{\banachfunspace}[1]{\setL^{#1}}	
\safemath{\hilfunspace}{\banachfunspace{2}}	
\safemath{\hilfunspacep}{\hilfunspace(\complexset)}	
\safemath{\schwarzspace}{\setS}				

\newcommand{\hadj}[1]{#1^{\star}}			

\safemath{\SNR}{\rho} 				
\safemath{\SINR}{\text{\sc sinr}} 				
\safemath{\No}{N_0}							
\safemath{\Es}{E_s}							
\safemath{\Eb}{E_b}							
\safemath{\EbNo}{\frac{\Eb}{\No}}
\safemath{\EsNo}{\frac{\Es}{\No}}
\safemath{\NoVar}{\variance}                 

\let\time\undefined
\safemath{\time}{\sct}						
\safemath{\dtime}{\sck}						
\safemath{\delay}{\sctau}					
\safemath{\ddelay}{\scl}						
\safemath{\doppler}{\scnu}					
\safemath{\ddoppler}{\scm}					
\safemath{\freq}{\scf}						
\safemath{\dfreq}{\scn}						
\safemath{\Dtime}{\Delta\time}
\safemath{\Dfreq}{\Delta\freq}
\safemath{\Ddtime}{\dtime}
\safemath{\Ddfreq}{\dfreq}
\safemath{\bandwidth}{\scB}
\safemath{\maxdoppler}{\doppler_{0}}			
\safemath{\maxdelay}{\delay_{0}}				
\safemath{\spread}{\Delta_{\CHop}}			
\DeclareMathOperator{\CHop}{\ensuremath{\opH}} 
\safemath{\kernel}{\rndk_{\CHop}}			
\safemath{\kernelp}{\kernel(\time,\time')}	
\safemath{\tvir}{\rndh_{\CHop}}				
\safemath{\tvirp}{\tvir(\time,\delay)}		
\safemath{\tvirc}{\conj{\rndh}_{\CHop}}		
\safemath{\tvtf}{\rndl_{\CHop}}				
\safemath{\tvtfp}{\tvtf(\time,\freq)}			
\safemath{\tvtfc}{\conj{\rndl}_{\CHop}}		
\safemath{\spf}{\rnds_{\CHop}}				
\safemath{\spfp}{\spf(\doppler,\delay)}		
\safemath{\spfc}{\conj{\rnds}_{\CHop}}		
\safemath{\bff}{\rndb_{\CHop}}				
\safemath{\bffp}{\bff(\doppler,\freq)}		

\safemath{\irc}{\scr_{\rndh}}				
\safemath{\tfc}{\scr_{\rndl}}				
\safemath{\spc}{\scr_{\rnds}}				
\safemath{\bfc}{\scr_{\rndb}}				

\safemath{\scaf}{\scc_{\rnds}}				
\safemath{\scafp}{\scaf(\doppler,\delay)}		
\safemath{\ccf}{\scc_{\rndl}}				
\safemath{\ccfp}{\ccf(\Dtime,\Dfreq)}			
\safemath{\cic}{\scc_{\rndh}}				
\safemath{\cicp}{\cic(\Dtime,\delay)}			

\safemath{\mi}{I}							
\safemath{\capacity}{C}					

\DeclareMathOperator{\Prob}{\opP}		

\safemath{\normal}{\mathcal{N}}			
\safemath{\jpg}{\mathcal{CN}}			
\safemath{\uniform}{\mathcal{U}}				
\safemath{\mchain}{\leftrightarrow}		

\safemath{\dB}{\,\mathrm{dB}}
\safemath{\dBm}{\,\mathrm{dBm}}
\safemath{\Hz}{\,\mathrm{Hz}}
\safemath{\kHz}{\,\mathrm{kHz}}
\safemath{\MHz}{\,\mathrm{MHz}}
\safemath{\GHz}{\,\mathrm{GHz}}
\safemath{\s}{\,\mathrm{s}}
\safemath{\ms}{\,\mathrm{ms}}
\safemath{\mus}{\,\mathrm{\text{\textmu}s}}
\safemath{\ns}{\,\mathrm{ns}}
\safemath{\ps}{\,\mathrm{ps}}
\safemath{\meter}{\,\mathrm{m}}
\safemath{\mm}{\,\mathrm{mm}}
\safemath{\cm}{\,\mathrm{cm}}
\safemath{\m}{\,\mathrm{m}}
\safemath{\W}{\,\mathrm{W}}
\safemath{\mW}{\, \mathrm{mW}}
\safemath{\J}{\,\mathrm{J}}
\safemath{\K}{\,\mathrm{K}}
\safemath{\bit}{\,\mathrm{bit}}
\safemath{\nat}{\,\mathrm{nat}}

\safemath{\define}{\triangleq}					

\safemath{\equivalent}{\sim}
\safemath{\distas}{\sim}					
\safemath{\sdiff}{\Delta}				
\safemath{\setdiff}{\setminus}				

\safemath{\reals}{\mathbb R}
\safemath{\positivereals}{\reals^{+}}
\safemath{\integers}{\mathbb Z}
\safemath{\posint}{\integers^{+}}
\safemath{\naturals}{\mathbb N}
\safemath{\posnaturals}{\naturals^{+}}
\safemath{\complexset}{\mathbb C}
\safemath{\rationals}{\mathbb Q}

\safemath{\iSet}{\setI}
\safemath{\rel}{\bowtie}					
\safemath{\eqrel}{\sim}					
\safemath{\rlord}{\leq}					
\safemath{\slord}{<}						
\safemath{\rpord}{\preceq}				
\safemath{\rrpord}{\succeq}				
\safemath{\spord}{\prec}					
\safemath{\sig}{\sigma}					
\safemath{\metric}{d}					

\safemath{\setfun}{\Phi}					
\safemath{\measure}{\mu}					
\safemath{\altmeasure}{\lambda}					
\newcommand{\outerm}[1]{#1^{\star}}		
\newcommand{\innerm}[1]{#1_{\star}}		
\safemath{\omeasure}{\outerm{\measure}}		
\safemath{\imeasure}{\innerm{\measure}}		
\safemath{\aecol}{\colS^{\star}_{\measure}} 
\safemath{\emeasure}{\bar{\measure}_{0}}	
\safemath{\rmeasure}{\tilde{\measure}}	
\safemath{\bmeasure}{\measure_{0}}		
\safemath{\glength}{\measure_{\altfun}}	
\safemath{\lebmea}{\lambda}				
\safemath{\blebmea}{\lebmea_{0}}			
\safemath{\sfun}{s}						
\safemath{\absintspace}{\colL^{1}}		
\safemath{\sqintspace}{\colL^{2}}		
\safemath{\abssumspace}{l^{1}}		
\safemath{\sqsumspace}{l^{2}}		

\safemath{\sfield}{\setF}				
\safemath{\vectors}{\setV}				
\safemath{\vecspace}{(\vectors,\sfield)}	
\safemath{\linspace}{\setV}				
\safemath{\clinspace}{(\linspace,\sfield)} 
\safemath{\nspace}{\setU}				
\safemath{\metspace}{\setM}				
\safemath{\bspace}{\setB}				
\safemath{\ipspace}{\setG}				
\safemath{\hilspace}{\setH}				
\safemath{\blospace}{\setG}				
\safemath{\lop}{\opT}					
\safemath{\altlop}{\opS}					
\safemath{\nullsp}{\nullspace(\lop)}		
\safemath{\lfun}{l}						
\safemath{\altlfun}{g}					
\newcommand{\dual}[1]{#1^{'}}			
\safemath{\dsum}{\oplus}					
\safemath{\funspace}{\colL}				
\renewcommand{\adj}[1]{#1^{\times}}		
\safemath{\adjlop}{\adj{\lop}}			
\safemath{\hadjlop}{\hadj{\lop}}			
\safemath{\tow}{\xrightarrow{w}}			
\safemath{\tows}{\xrightarrow{w^{*}}}		
\safemath{\cparam}{\lambda}
\safemath{\lopl}{\lop_{\cparam}}		
\safemath{\iop}{\opI}					
\safemath{\resolop}{\opR}				
\safemath{\resolvent}{\resolop_{\cparam}(\lop)}	
\safemath{\reset}{\setQ}
\safemath{\spectrum}{\setS}
\safemath{\resolset}{\reset(\lop)}		
\safemath{\lopspec}{\spectrum(\lop)}		
\safemath{\pspec}{\spectrum_{p}(\lop)}	
\safemath{\cspec}{\spectrum_{c}(\lop)}	
\safemath{\rspec}{\spectrum_{r}(\lop)}	
\safemath{\ev}{\cparam}					
\newcommand{\specrad}[1]{r_{#1}}			
\safemath{\lopsrad}{\specrad{\lop}}		
\safemath{\pop}{\opP}					

\safemath{\specfam}{\colE}				
\safemath{\specop}{\opE_{\cparam}}		
\safemath{\altspecop}{\opE_{\mu}}		
\safemath{\vmulti}{\vecone}				
\safemath{\unitaryop}{\opU}				
\safemath{\sval}{\sigma}					
\safemath{\corrcoef}{\rho}				
\safemath{\sangle}{\theta}				

\let\time\undefined
\safemath{\iset}{\setI}				
\safemath{\shift}{\nu}
\safemath{\scale}{\alpha}
\safemath{\time}{t}
\safemath{\specfreq}{\theta}	
\newcommand{\transopgen}[1]{\opT_{#1}} 
\safemath{\transop}{\transopgen{\delay}}
\newcommand{\modopgen}[1]{\opM_{#1}}	
\safemath{\modop}{\modopgen{\shift}}
\newcommand{\dilopgen}[1]{\opD_{#1}}	
\safemath{\dilop}{\dilopgen{\scale}}
\safemath{\fram}{\setF}				
\safemath{\dfram}{\dual{\fram}}		
\safemath{\ufb}{B}					
\safemath{\lfb}{A}					
\safemath{\sop}{\hadj{\aop}}				
\safemath{\aop}{\opT}			
\safemath{\fop}{\opS}				
\safemath{\daop}{\tilde\opT}			
\safemath{\dsop}{\hadj{\tilde\opT}}				

\safemath{\ifop}{\inv{\fop}}			
\safemath{\rifop}{\fop^{-1/2}}			
\safemath{\transeq}{\setT}			
\safemath{\nfun}{\Phi}				
\safemath{\funvec}{\vecf}			
\safemath{\altfunvec}{\vecg}

\safemath{\samplespace}{\Omega}
\safemath{\probspace}{(\samplespace,\sfield,\Prob)}	
\safemath{\ccoef}{\rho}			

\safemath{\infstate}{\vecpi}				
\safemath{\typset}{\setA_{\epsilon}^{(N)}}	
\safemath{\expequal}{\doteq}				
\safemath{\code}{C}						
\safemath{\dstringset}{\setD^{\star}}		
\safemath{\cwlength}{l}					
\safemath{\elength}{L}					
\safemath{\extension}{C^{\star}}			
\safemath{\approaches}{\rightarrow}		
\safemath{\evnt}{\setA}					
\safemath{\altevnt}{\setB}					
\safemath{\rv}{\rndx}					
\safemath{\altrv}{\rndy}					
\safemath{\complexrv}{\rndu}					
\safemath{\altcrv}{\rndv}				
\safemath{\rvec}{\rvecx}					
\safemath{\altrvec}{\rvecy}				
\safemath{\crvec}{\rvecu}				
\safemath{\altcrvec}{\rvecv}				
\safemath{\variance}{\sigma^{2}}			
\safemath{\map}{T}						
\safemath{\jacobian}{\matJ}					
\safemath{\wvec}{\rvecw}					
\safemath{\wrv}{\rndw}					
\safemath{\orthmat}{\matQ}				
\safemath{\evmat}{\matLambda}			
\safemath{\identity}{\matidentity}		
\safemath{\innovec}{\vecv}				
\safemath{\convas}{\xrightarrow{\text{a.s.}}}	
\safemath{\convr}{\xrightarrow{\text{r}}}	
\safemath{\convp}{\xrightarrow{\text{P}}}	
\safemath{\convd}{\xrightarrow{\text{D}}}	
\safemath{\ltis}{\opL}				
\safemath{\ir}{h}					
\safemath{\tf}{\MakeUppercase{\ir}}	

\newcommand*{\fancyrefparlabelprefix}{par}		
\newcommand*{\fancyrefremlabelprefix}{rem}		
\newcommand*{\fancyrefchalabelprefix}{cha}		
\newcommand*{\fancyrefapplabelprefix}{app}		
\newcommand*{\fancyrefthmlabelprefix}{thm}		
\newcommand*{\fancyreflemlabelprefix}{lem}		
\newcommand*{\fancyrefcorlabelprefix}{cor}		
\newcommand*{\fancyrefdeflabelprefix}{def}		
\newcommand*{\fancyrefproplabelprefix}{prop}		
\frefformat{vario}{\fancyrefparlabelprefix}{Part~#1}
\frefformat{vario}{\fancyrefremlabelprefix}{Remark~#1}
\frefformat{vario}{\fancyrefchalabelprefix}{Chapter~#1}
\frefformat{vario}{\fancyrefseclabelprefix}{Section~#1}

\frefformat{vario}{\fancyrefthmlabelprefix}{Theorem~#1}
\frefformat{vario}{\fancyreflemlabelprefix}{Lemma~#1}
\frefformat{vario}{\fancyrefcorlabelprefix}{Corollary~#1}
\frefformat{vario}{\fancyrefdeflabelprefix}{Definition~#1}
\frefformat{vario}{\fancyreffiglabelprefix}{Figure~#1}
\frefformat{vario}{\fancyrefapplabelprefix}{Appendix~#1}
\frefformat{vario}{\fancyrefeqlabelprefix}{(#1)}
\frefformat{vario}{\fancyrefproplabelprefix}{Property~#1}

\theoremstyle{definition}
\newtheorem{thm}{Theorem}
\newtheorem{lem}{Lemma}

\newtheorem{cor}{Corollary}

\newtheorem{exa}{Example}
\newtheorem{dfn}{Definition}

\usepackage{pgfplots}
\pgfplotsset{compat=1.14}
\usepackage{enumitem}
\usepackage{xr} 		
\allowdisplaybreaks[3]
\mathtoolsset{showonlyrefs=true}   
\usetikzlibrary{cd,matrix,patterns,calc,decorations.fractals,arrows.meta,arrows,external,pgfplots.colormaps,lindenmayersystems}
\definecolor{tblblue}{rgb}{0.93,0.93,1.0}
\definecolor{tblred}{rgb}{1,0.93,0.93}
\definecolor{darkblue}{rgb}{0,0,0.7} 
\definecolor{darkgreen}{RGB}{20,120,43} 
\definecolor{darkred}{rgb}{0.8,0,0} 
\definecolor{lightblue}{RGB}{101,124,191}
\definecolor{skyblue}{RGB}{135,206,235}
\definecolor{gold}{RGB}{204,168,66}
\definecolor{strongblue}{RGB}{60,146,228}
\definecolor{lightgray}{gray}{0.5}
\definecolor{verylightgray}{RGB}{101,124,191}
\definecolor{mistyrose}{RGB}{238,213,210}
\definecolor{firebrick3}{RGB}{205,38,38}
\pgfdeclarelindenmayersystem{cantor set}{
  \rule{F -> FfF}
  \rule{f -> fff}
}

 \author{\IEEEauthorblockN{
Erwin Riegler$^1$, 
G\"unther Koliander$^2$, 
and
Helmut B\"olcskei$^1$\\ 
\medskip}
\IEEEauthorblockA{$^1$Dept.~IT~\&~EE, ETH Zurich, Zurich, Switzerland, 
Email: \{eriegler, boelcskei\}@nari.ee.ethz.ch}
\IEEEauthorblockA{$^2$Acoustics Research Institute, Austrian Academy of Sciences, Vienna,  Austria,
Email: 	gkoliander@kfs.oeaw.ac.at}
}

\begin{document}
\title{Rate-Distortion Theory for General\\[-1mm] Sets and Measures}
\maketitle
\begin{abstract}
This paper is concerned with a rate-distortion theory for  sequences of i.i.d. random variables with general distribution supported on general sets including manifolds and fractal sets. Manifold structures are prevalent in data science, e.g., in compressed sensing, machine learning, image processing, and handwritten digit recognition. Fractal sets find application in image compression and in modeling of Ethernet traffic. We derive a lower bound  on the (single-letter) rate-distortion function that applies to random variables $X$ of general distribution $\mu_X$ and for continuous $X$ reduces to the classical Shannon lower bound. Moreover, our lower bound is explicit up to a parameter obtained by solving a convex optimization problem in a nonnegative real variable. The only requirement for the bound to apply is the existence of a $\sigma$-finite reference measure $\mu$ for $X$ (i.e., a measure $\mu$ with $\mu_X\ll\mu$ and such that the generalized entropy $h_\mu(X)$ is finite) satisfying a certain subregularity condition. This condition is very general and prevents the reference measure $\mu$ from being highly concentrated on balls of small radii. To illustrate the wide applicability of our result, we evaluate the lower bound for a random variable distributed uniformly on a manifold, namely, the unit circle, and a random variable distributed uniformly on a self-similar set, namely, the middle third Cantor set. 
\end{abstract}

\section{Introduction and Mathematical Setup} 
%
%
This paper is concerned with a rate-distortion (R-D) theory for sequences of i.i.d. random variables with general distribution supported on general sets including manifolds and fractal sets. 
Manifold structures are prevalent in data science, e.g., in compressed sensing \cite{bawa09,care09,capl11,albdekori18,ristbo15}, machine learning \cite{lz12}, image processing \cite{lufahe98,soze98}, and handwritten digit recognition \cite{hidare97}. 
Fractal sets find application in image compression and in modeling of Ethernet traffic \cite{letawi94}.

R-D theory \cite{sh59,be71,gr90,grne98} is concerned with the characterization of ultimate limits on the discretization of sequences of random variables. 
Specifically, let  $(\setX,\colX)$ and $(\setY,\colY)$ be measurable spaces equipped with a measurable function  $\rho\colon \setX\times\setY\to[0,\infty]$, henceforth called  distortion function, and let $(X_i)_{i\in\naturals}$ be a sequence of random variables  with the $X_i$  distributed on $\setX$. 
For every $l\in\naturals$, one considers all measurable mappings $g_l\colon \setX^l \to \setY^l$ with  $\lvert g_l(\setX^l) \rvert <\infty$, referred to as source codes of length $l$.  
A pair $(R,D)$ of nonnegative real numbers is said to be {achievable} if, for sufficiently large $l\in\naturals$, 
there exists a source code $g_l$ of length $l$ with $|g_l(\setX^l)| \leq \lfloor e^{lR}\rfloor$ and  expected average distortion  
\begin{align}\label{eq:leqd}
\opE\mleft[\frac{1}{l}\sum_{i=1}^l\rho(X_i,(g_l(X_1,\dots,X_l))_i)\mright] \leq D .
\end{align} 
Suppose that  $(\setX,\colX)$ and $(\setY,\colY)$ are standard spaces  
(cf. \cite[Section 1.4]{gr11}) and consider a sequence $(X_i)_{i\in\naturals}$ of  i.i.d. random variables that are distributed on $\setX$. The  
(single-letter) R-D function is defined as   
\begin{align}\label{eq:RD}
R(D):=
\inf_{Y:\,\opE[\rho(X,Y)]\,\leq\, D} I(X,Y),  
\end{align}
where  $Y$ is distributed on $(\setY,\colY)$, $X=X_1$, and $I(\cdot,\cdot)$ denotes mutual information. 
If there exists a  $y^*\in\setY$ with  
$\opE\mleft[\rho(X,y^*)\mright]<\infty$,  
then the R-D theorem \cite[Theorems 7.2.4 \& 7.2.5]{be71} states that
\begin{enumerate}
\renewcommand{\theenumi}{\roman{enumi})}
\renewcommand{\labelenumi}{\roman{enumi})}
\item
for every $D\geq 0$ with $R(D)<\infty$, $(R,D)$ is achievable for all $R>R(D)$, and 
\item  $(R,D)$ is not achievable for all $R<R(D)$.
\end{enumerate}

The  function $ R(D)$ is difficult to characterize analytically in general, but 
asymptotic results in terms of the R-D dimension of order $k>0$, defined as 
$-(1/k)\lim_{D\to 0} R(D)/\log D$  if the limit exists, are available \cite{kade94}.  
For discrete-continuous mixtures, the function  $ R(D)$ is known explicitly up to a term that vanishes as  $D\to 0$ \cite{ro88}.  
For general distributions, only bounds on  $ R(D)$ are available. 
While upper bounds on $ R(D)$ can be obtained by evaluating  $I(X,Y)$ for a specific $Y$   with $\opE[\rho(X,Y)]\leq D$, 
lower bounds  are notoriously hard to obtain. 
The best-known lower bound is the Shannon lower bound for  discrete random variables of finite entropy and with $\sum_{x\in\setX}e^{-s\rho(x,y)}$ independent of $y$ for all $s>0$ \cite[Section 4.3]{gr90}, and for continuous random variables of finite differential entropy and with difference distortion function $\rho(x-y)$ \cite[Section 4.6]{gr90}. 
For continuous $X$ of finite differential entropy and distortion function $\rho(x-y)=\lVert x-y\rVert_\text{s}^k$, where $\lVert\,\cdot\,\rVert_\text{s}$ is a semi-norm and $k>0$, the Shannon lower bound is known explicitly \cite[Section VI]{yatagr80} and, 
provided that $X$ satisfies a certain moment constraint, tight as $D\to 0$ \cite{liza94,ko16}. 
Using Csisz\'ar's parametric representation of  $ R(D)$  \cite{cs74}, a Shannon lower  bound   was reported recently in \cite[Theorem 55]{kopirihl16} for the class of  $m$-rectifiable random variables \cite[Definition 11]{kopirihl16}, and  
for  general random variables in  \cite[Theorem 2]{ko17}.  
The bounds in \cite{kopirihl16,ko17} are, however, not explicit.  
 
\emph{Contributions.}  
We derive  a lower bound $R_{\text{L}}(X)$ on  the {R-D} function $ R(D)$ in \eqref{eq:RD} 
 for  random variables  $X$  of general distribution supported  on general sets  including manifolds and fractal sets. 
The expression for $R_{\text{L}}(X)$ we get is explicit up to a parameter obtained by solving a convex optimization problem in a nonnegative real variable and, for continuous $X$ of finite differential entropy and distortion function $\rho(x-y)=\lVert x-y\rVert_\text{s}^k$, reduces to the classical Shannon lower bound reported in   \cite{yatagr80}.   
The only requirement for  our lower bound to apply is the existence of a $\sigma$-finite reference measure $\mu$ for $X$ (i.e., a measure $\mu$ with $\mu_X\ll\mu$ and such that the generalized entropy $h_\mu(X)$ is finite)  satisfying a certain subregularity condition.  
This subregularity condition guarantees the existence of a $\delta_0>0$ such that the reference measure $\mu$ is not  highly concentrated on balls of radii $\delta\in(0,\delta_0]$;  it is satisfied, e.g., by uniform distributions 
on  regular sets of dimension $m$ in $\reals^d$ (cf. \cite[Section 12]{grlu00}). 
Specific examples of regular sets of dimension $m$ are compact convex sets $\setK\subseteq\reals^m$ with $\operatorname{span}(\setK)=\reals^m$ \cite[Example 12.7]{grlu00}, surfaces of compact convex sets $\setK\subseteq\reals^{m+1}$ with $\operatorname{span}(\setK)=\reals^{m+1}$ \cite[Example 12.8]{grlu00}, 
$m$-dimensional compact $C^1$-submanifolds of $\reals^d$ \cite[Example 12.9]{grlu00},  
self-similar sets of similarity dimension $m$ satisfying the weak separation property \cite[Theorem 2.1]{frheolro15}, and finite unions 
of regular sets of dimension $m$ \cite[Lemma 12.4]{grlu00}. 
To illustrate the wide applicability of our result, we evaluate the lower bound $R_{\text{L}}(X)$ for a random variable distributed uniformly on a manifold, namely, the unit circle, and for a random variable distributed uniformly on a self-similar set, namely, the middle third Cantor set. 
Proofs are omitted throughout due to space constraints. 

\emph{Notation.} 
Sets are designated by calligraphic letters, e.g., $\setA$, with $|\setA|$ denoting  cardinality and   
$\overline{\setA}$  closure.   
$\sigma$-algebras are indicated by script letters, e.g., $\colX$,  
and will throughout be assumed to contain all singleton sets.  
For a measure space $(\setX,\colX,\mu)$ and a measurable set $\setA \in \colX$, we write $\mu|_{\setA}$ for the restriction of $\mu$ to $\setA$. 
For a Borel measure $\mu$, the support  $\operatorname{supp}(\mu)$ is the smallest closed set such that $\mu(\setX\mysetminus \operatorname{supp}(\mu))=0$. 
We denote the $m$-dimensional Hausdorff measure  by $\colH^m$ \cite[Definition 2.46]{amfupa00}. 
For $\mu$ and $\nu$ defined on the same measurable space with 
$\mu$ absolutely  continuous with respect to $\nu$, expressed by  $\mu\ll\nu$,  we write  $\mathrm d\mu/\mathrm d \nu$ for  the Radon-Nikodym derivative of $\mu$ with respect to $\nu$.  
The product measure of $\mu$ and $\nu$ is designated by $\mu\otimes\nu$. 
Random variables distributed on general measurable spaces $(\setX,\colX)$  are denoted by  capital letters, e.g., $X$, and $\mu_X$ is the distribution of  $X$. 
$\opE[\cdot]$ stands for the expectation operator.  
If $X$ is distributed on the $\sigma$-finite measure space $(\setX,\colX,\mu)$ and of finite  generalized entropy  
\begin{align}
h_\mu(X)
&:=-\opE\mleft[\log \frac{\mathrm d\mu_X}{\mathrm d\mu}(X)\mright]  
\end{align}
with $\mu_X\ll\mu$, then we call $\mu$ a reference measure for $X$.  
For $X$ distributed on $(\setX,\colX)$  and $Y$ distributed on $(\setY,\colY)$, the mutual information between $X$ and $Y$ is 
\begin{align}
I(X,Y):=\opE\mleft[\log \frac{\mathrm d\mu_{X,Y}}{\mathrm d(\mu_X\otimes\mu_Y)}(X,Y)\mright]
\end{align}
if $\mu_{X,Y}\ll\mu_X\otimes\mu_Y$, and $I(X,Y):=\infty$ else.  
For $a>0$, the gamma function is defined by $\Gamma(a)=\int_0^\infty t^{a-1}e^{-t}\,\mathrm d t$.  
For $a>0$ and $s\geq 0$, the lower incomplete gamma function is $\gamma(a,s)=\int_0^s t^{a-1}e^{-t}\,\mathrm d t$. 
Norms on $\reals^d$ are denoted as $\lVert\,\cdot\,\rVert$,   $\lVert\,\cdot\,\rVert_2$ stands for the  Euclidean norm, and $\lVert\,\cdot\,\rVert_\text{s}$ refers to a general semi-norm.  
For $a\in \reals$, we let  $\lfloor a \rfloor$ be  the greatest integer less than or equal to $a$.   
For $a>0$, $\log a$ denotes the  logarithm of $a$ taken to the base $e$.  
We use the convention  $0\cdot\infty=0$.  

\section{The Subregularity Condition} 
Our  lower bound on the R-D function is valid for reference measures $\mu$ satisfying the following  subregularity condition, which  prevents  $\mu$ from being highly concentrated on balls of small radii. 
\begin{dfn}\label{def:subreg}
Let $(\setX,\colX,\mu)$ be a 
measure space, $(\setY,\colY)$ a measurable space,  $\rho\colon \setX\times\setY\to[0,\infty]$ 
a   distortion function,  $k>0$, and set $\setB_{\rho^{1/k}}\mleft(y,\delta\mright):=\{x\in\setX:\rho^{1/k}(x,y)<\delta\}$.  
The  measure $\mu$ is {$\rho^{1/k}$-subregular of dimension $m$} if there exist 
constants $\delta_0\in(0,\infty]$ and $c>0$ such that  
\begin{align}\label{eq:subregularity}
\mu\mleft(\setB_{\rho^{1/k}}\mleft(y,\delta\mright)\mright)\leq c\delta^m\quad\text{for all $y\in\setY$ and  $\delta\in (0,\delta_0)$}.
\end{align}
The  measure $\mu$ is {$\rho^{1/k}$-regular of dimension $m$} if there exist 
constants $\delta_0\in(0,\infty]$ and $c^\prime,c>0$ such that  
\begin{align}\label{eq:regularity}
c^\prime\delta^m
\leq \mu\mleft(\setB_{\rho^{1/k}}\mleft(y,\delta\mright)\mright)
\leq c\delta^m\quad\!\text{for all $y\in\setY$ and $\delta\in (0,\delta_0)$}. 
\end{align} 
\end{dfn} 
Lebesgue measure on $\setX=\setY=\reals^d$ together with $\rho(x,y)=\lVert x-y\rVert_\mathrm{s}^k$ satisfies \eqref{eq:regularity} with  $c^\prime=c$.  Discrete measures do not satisfy \eqref{eq:subregularity}. 
For the particular choices $\setX=\reals^d$, $\lVert\,\cdot\,\rVert$ a norm on $\reals^d$, $\mu$ a Borel measure, and $\setY=\operatorname{supp}(\mu)$, $\lVert\,\cdot\,\rVert$-regularity of dimension $m$ agrees with regularity of dimension $m$ as introduced in \cite[Definition 12.1]{grlu00}. 
A compact set $\setK\subseteq \reals^d$ with   $0<\colH^m(\setK)<\infty$ is called regular of dimension $m$
if the measure $\colH^m|_\setK$ is $\lVert\,\cdot\,\rVert$-regular (and hence also $\lVert\,\cdot\,\rVert$-subregular) of dimension $m$ \cite[Definition 12.1]{grlu00}. 
Specific examples of regular sets of dimension $m$ are 
compact convex sets $\setK\subseteq\reals^m$ with $\operatorname{span}(\setK)=\reals^m$ \cite[Example 12.7]{grlu00}, surfaces of compact convex sets $\setK\subseteq\reals^{m+1}$ with $\operatorname{span}(\setK)=\reals^{m+1}$ \cite[Example 12.8]{grlu00}, 
$m$-dimensional compact $C^1$-submanifolds of $\reals^d$ \cite[Example 12.8]{grlu00},  
self-similar sets of similarity dimension $m$ satisfying the weak separation property \cite[Theorem 2.1]{frheolro15}, and finite unions 
of regular sets of dimension $m$ \cite[Lemma 12.4]{grlu00}. 

If  $\mu(\setX)<\infty$ and the subregularity condition  \eqref{eq:subregularity} holds for some $c,\delta_0>0$, then $c$ can be modified to make \eqref{eq:subregularity} hold  for $\delta_0=\infty$. 
The formal statement is as follows.

\begin{lem}\label{lem:rho}
Let $(\setX,\colX,\mu)$ be a measure space with $\mu(\setX)<\infty$, $(\setY,\colY)$ a measurable space, 
$\rho\colon \setX\times\setY\to[0,\infty]$ 
a   distortion function,  and $k>0$. 
If there exist constants  $c,\delta_0>0$ such that $\mu$ satisfies the subregularity condition \eqref{eq:subregularity}, then 
\begin{align}\label{eq:global}
\mu\mleft(\setB_{\rho^{1/k}}\mleft(y,\delta\mright)\mright)\leq 
\max(c,\mu(\setX)\delta_0^{-m})\delta^m 
\end{align}
for all $y\in\setY$ and  $\delta>0$.
\end{lem}

\section{Lower Bound on the Rate-Distortion Function} \label{sec:slb}
Based on the parametric representation of $R(D)$  in \cite[Theorem 2.3]{cs74},     
a  Shannon lower bound for rectifiable measures \cite[Definition 2.59]{amfupa00} as reference measures was reported recently in \cite[Theorem 55]{kopirihl16}.  
We now extend this bound to general (not necessarily rectifiable) reference measures $\mu$. 
\begin{lem}\label{thm.ko17}
Consider a random variable  $X$ distributed on the measure space $(\setX,\colX,\mu)$,  a measurable space $(\setY,\colY)$, and  a distortion function  $\rho\colon \setX\times\setY\to[0,\infty]$ satisfying  
\begin{enumerate}
\renewcommand{\theenumi}{\roman{enumi})}
\renewcommand{\labelenumi}{\roman{enumi})}
\item $\inf_{y\in\setY}\rho(x,y)=0$ for all $x\in\setX$, and \label{cond:infzero}
\item there exists a finite set $\setB\subseteq\setY$ such that   
$
\opE\mleft[\min_{y\in\setB}\rho(X,y)\mright]<\infty 
$. \label{cond:refpoint}
\end{enumerate}
Suppose that  $\mu$ is a reference measure for $X$ and 
let $D_0:=\inf\{D\geq 0:R(D)<\infty\}$. 
Then, $R(D)\geq R_{\text{SLB}}(D)$ for all $D\in(D_0,\infty)$, where 
\begin{align}
R_{\text{SLB}}(D)= h_\mu(X)-\inf_{s\geq 0}\mleft(sD +\log \nu(s)\mright)\label{eq:SLB}
\end{align}
with 
\be
\nu(s)=\sup_{y\in\setY}\int e^{-s\rho(x,y)} \mathrm d \mu(x). \label{eq:defns1}
\ee
\end{lem}

For discrete $X$   of finite entropy, $\mu$  the counting measure,  and $\sum_{x\in\setX}e^{-s\rho(x,y)}$ independent of $y$ for all $s>0$,  Lemma \ref{thm.ko17} recovers the  Shannon lower bound for discrete random variables reported in \cite[Lemma 4.3.1]{gr90}. 
For $X$ continuous, $\mu$ the Lebesgue measure, $\setX=\setY=\reals^d$, and $\rho(x,y)=\rho(x-y)$,   Lemma \ref{thm.ko17} recovers  the Shannon lower bound for continuous random variables \cite[Equation 4.6.1]{gr90}, which  
can be evaluated explicitly for $\rho(x,y)=\lVert x-y\rVert^k_\mathrm s$ with $k>0$, leading to the classical form of the Shannon lower bound \cite[Section VI]{yatagr80}
\be \label{eq:shlbclassic}
R_{\text{SLB}}(D) = 
h(X) +\log\mleft(\frac{\mleft(\frac{d}{kD}\mright)^\frac{d}{k}}{V_d\, \Gamma\mleft(\frac{d}{k}+1\mright)}\mright)-\frac{d}{k}.
\ee
Here, $V_d$ is the Lebesgue measure of the unit ball with respect to the semi-norm $\lVert\,\cdot\,\rVert_\text{s}$. 
What makes the explicit expression \eqref{eq:shlbclassic} possible is the following 
simplification of $\nu(s)$ in \eqref{eq:defns1} for difference distortion functions $\rho(x,y)=\rho(x-y)$ and  translation invariant reference measures $\mu$, namely 
\begin{align}
\nu(s)
&= \sup_{y\in\setY}\int e^{-s\rho(x-y)} \,\mathrm d \mu(x)\\
&= \int e^{-s\rho(x)} \,\mathrm d \mu(x),\label{eq:defns1b}
\end{align}
which can be evaluated explicitly for $\rho(x,y)=\lVert x-y\rVert^k_\mathrm s$ with $k>0$ by changing  variables to polar coordinates. 
Unfortunately, for $X$ of general distribution and for general distortion functions, $\nu(s)$ in \eqref{eq:defns1} cannot be further simplified, which precludes  an explicit expression for $R_{\text{SLB}}(D)$.    
However, if the  reference measure $\mu$ is $\rho^{1/k}$-subregular, then we can 
 upper-bound  $\nu(s)$. This leads to  a  lower bound on $R(D)$  that is explicit 
 up to a parameter  obtained by solving a  
  convex optimization problem in a nonnegative real variable.
 The corresponding formal statement is as follows. 

\begin{thm}\label{thm:new}
Consider a random variable  $X$ distributed on the measure space $(\setX,\colX,\mu)$,  a measurable space $(\setY,\colY)$, and  a distortion function  $\rho\colon \setX\times\setY\to[0,\infty]$ satisfying  
Properties \ref{cond:infzero} and \ref{cond:refpoint} stated in Lemma  \ref{thm.ko17}. 
Suppose that  $\mu$ is a $\rho^{1/k}$-subregular reference measure for $X$  of dimension $m$ satisfying \eqref{eq:subregularity} with  
$\delta_0\in(0,\infty]$ and $c>0$,   
and 
let $D_0:=\inf\{D\geq 0:R(D)<\infty\}$. 
Suppose further that either $\delta_0=\infty$ or $\mu(\setX)<\infty$. 
Then,  
\begin{align}\label{eq:toshow1}
R_{\text{SLB}}(D)&\geq R_{\text{L}}(D)\quad\text{for all $D>D_0$},    
\end{align}
where $R_{\text{L}}(D)$ is given by
\begin{align}
&R_{\text{L}}(D) =\\ 
&\begin{cases}
h_\mu(X) +\log\mleft( \frac{\mleft(\frac{m}{kD} \mright)^\frac{m}{k}}{c\,\Gamma\mleft(\frac{m}{k}+1\mright)}\mright) -\frac{m}{k}
&\quad\text{if}\ c\geq \mu(\setX)\delta_0^{-m}\\
h_\mu(X) -\min_{s\geq 0}q(s,D)&\quad\text{else}, 
\end{cases}\label{eq:SLB1}
\end{align}
where 
\be
q(s,D)=    s\delta_0^{-k}D + p(s)\label{eq:qs}
\ee
with
\be
p(s)=\log\mleft(\frac{\mu(\setX)\Gamma\mleft(\frac{m}{k}+1\mright)-\mleft(\mu(\setX)-\delta_0^m c\mright)\gamma\mleft(\frac{m}{k}+1,s\mright)}{s^\frac{m}{k}}\mright).\label{eq:ks} 
\ee 
For every $D>0$, the function $q(\cdot,D)$ is strictly convex  on $\reals_+$ and  attains its unique  minimum  at $s_0$ defined (implicitly) through $\delta_0^{k}\,p^\prime(s_0)=-D$. 
\end{thm}
The  lower bound $R_{\text{L}}(D)$ in \eqref{eq:SLB1} is  explicit in the regime $c\geq \mu(\setX)\delta_0^{-m}$; for $c<\mu(\setX)\delta_0^{-m}$, it is explicit up to a parameter obtained by solving a convex optimization problem in a nonnegative real variable. 
As the lower bound $R_{\text{L}}(D)$ is obtained from $R_{\text{SLB}}(D)$ in \eqref{eq:SLB} by upper-bounding $\nu(s)$ in \eqref{eq:defns1} making use of subregularity of the reference measure $\mu$,  it follows that $R_{\text{L}}(D)=R_{\text{SLB}}(D)$ whenever the reference measure satisfies the subregularity condition with equality and for $\delta_0=\infty$. 
Specifically, we have equality in the following special case. 
%
\begin{cor}
Consider a continuous random variable  $X$ distributed on $\reals^d$ and of finite differential entropy. 
Suppose that   $\rho(x,y)=\lVert x-y\rVert_\text{s}^k$ with $k>0$. Then,  $R_{\text{L}}(D)=R_{\text{SLB}}(D)$ for all $D\geq D_0$. 
\end{cor}

\section{Examples}

To illustrate the generality  of Theorem \ref{thm:new}, we consider two specific examples of random variables, 
namely a random variable  distributed uniformly on a manifold, specifically the unit circle, and a random variable distributed uniformly on a self-similar set, specifically the middle third Cantor set. 

\begin{exa}(Uniform distribution on the unit circle)\label{ex:S1A}
Let $\setX=\setY=\reals^2$ be equipped with the Borel $\sigma$-algebra and the distortion function $\rho(x,y)=\lVert x-y\rVert^2_2$,  and take $X$ 
 distributed uniformly on the unit circle $\setS_1\subseteq\reals^2$, 
i.e., $\mu_X=\colH^m|_{\setS_1}/\colH^m(\setS_1)$. 
We first  establish the subregularity condition  \eqref{eq:subregularity} for $\mu=\mu_X$, $k=2$, and $m=1$. 
It turns out that  (cf. Figure \ref{fig:S1a})
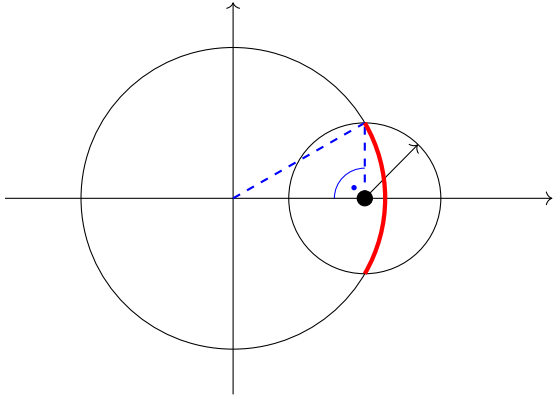
\begin{figure}[tb]
\begin{center}
\begin{tikzpicture}[scale=2]
\newcommand*{\rechterWinkel}[3]{
   \draw[shift={(#2:#3)}] (#1) arc[start angle=#2, delta angle=90, radius = #3];
   \fill[shift={(#2+45:#3/2)}] (#1) circle[radius=1.25\pgflinewidth];
}
\draw (0,0) circle(1);
\draw ({sqrt(1-0.5^2)},0) circle(0.5); 
\draw[arrows=->](0,-1.3)--(0,1.3); 
\draw[arrows=->](-1.5,0)--(2.1,0); 
\draw[thick,blue,dashed]({sqrt(1-0.5^2)},0)--({sqrt(1-0.5^2)},0.5);
\draw[thick,blue,dashed](0,0)--({sqrt(1-0.5^2)},0.5);
\draw[arrows=->]({sqrt(1-0.5^2)},0)--({sqrt(1-0.5^2)+0.35},{sqrt(0.5^2-0.35^2)});
\draw[ultra thick, red]  (1,0) arc (0:30:1);
\draw[ultra thick, red]  (1,0) arc (0:-30:1);
\filldraw[color=black] ({sqrt(1-0.5^2)},0) circle(0.05);
\put (42,-8){$x$};
\put (72,22){$\delta$};
\put (58,-15){\color{red}$\alpha$};
\put (-48,48){$\setS_1$};
\color{blue}
\rechterWinkel{{sqrt(1-0.5^2)},0}{90}{.2}
\end{tikzpicture}
\caption{For fixed  $\delta<1$, the maximum Hausdorff measure of the arc  $\alpha(x,\delta)=\setS_1\cap\setB_{\lVert\,\cdot\,\rVert_2}(x,\delta)$ is $\colH^1(\alpha(x,\delta))=2\arcsin(\delta)$, which is achieved for any $x\in \reals^2$ satisfying $\lVert x\rVert_2=\sqrt{(1-\delta^2)}$. 
\label{fig:S1a}}
\end{center}
\end{figure}
\begin{align}
\mu_X\mleft(\setB_{\lVert\,\cdot\,\rVert_2}\big(x,\delta \big)\mright)\label{eq:subpre1}
&=\mu_X(\{y\in\reals^2:\lVert y-x\rVert_2\leq \delta\})\\
&=\frac{\colH^1(\{y\in\setS_1:\lVert y-x\rVert_2\leq \delta\})}{2\pi}\\
&\leq \frac{\arcsin(\delta)}{\pi}\label{eq:subpre2}
\end{align}
for all  $\delta\in(0,1]$ and $x\in\reals^2$. 
Since  $\arcsin(x)/x$ is monotonically increasing  on $(0,1)$,  
 we can upper-bound  
$\arcsin(\delta)\leq \delta \frac{\arcsin(\hat{\delta})}{\hat{\delta}}$ for all  $\delta\in(0,\hat{\delta})$ and $\hat{\delta}\in (0,1]$.  
Therefore,  \eqref{eq:subpre1}--\eqref{eq:subpre2} leads to the family of subregularity conditions   
\begin{align}\label{eq:measupball}
\mu_X\mleft(\setB_{\lVert\,\cdot\,\rVert_2}\big(x,\delta\big)\mright)
&\leq  \frac{\arcsin(\hat{\delta})}{\pi\hat{\delta}}  \delta
\end{align}
for all $x\in\reals^2$ and $\delta\in(0,\hat\delta)$,  parametrized by $\hat{\delta}\in (0,1]$. 
For $\mu=\mu_X$, $m=1$, $k=2$, $\delta_0=\hat \delta\in(0,1]$, and $c=\arcsin(\hat \delta)/(\pi\hat\delta)$  
and hence $c< \mu_X(\setX)/\delta_0=1/\delta_0$,  the lower bound in \eqref{eq:SLB1} is given by   
\begin{align}
&R^{(\hat\delta)}_{\text{L}}(D):=\\ 
&-\frac{s_0}{\hat\delta^{2}}D 
-\log\mleft(\Gamma\mleft(\frac{3}{2}\mright)-\mleft(1-\frac{\arcsin(\hat \delta)}{\pi}\mright)\gamma\mleft(\frac{3}{2},s_0\mright)\mright)\\
&+\frac{1}{2}\log s_0\quad\text{for all $D>0$},
\end{align}
where $s_0$ is the unique solution of 
\begin{align} 
\frac{\hat\delta^2}{2s_0}+\frac{\hat \delta^2s_0^\frac{1}{2}e^{-s_0}}{\frac{\Gamma\mleft(\frac{3}{2}\mright)}{1-\frac{\arcsin(\hat \delta)}{\pi}}-\gamma\mleft(\frac{3}{2},s_0\mright)}=D.
\end{align}
Finally, we set 
\be\label{eq:SLB1S1}
R_{\text{L}}(D)=\max_{\hat\delta\in(0,1]}R^{(\hat\delta)}_{\text{L}}(D)\quad\text{for all $D>0$}.
\ee   
The result of the maximization in \eqref{eq:SLB1S1} carried out numerically is depicted in Figure~\ref{fig.bounds} along with the numerically evaluated Shannon lower bound $R_{\text{SLB}}(D)$ in \eqref{eq:SLB} from   \cite[Section X.C]{kopirihl16}. It can be seen that 
$R_{\text{L}}(D)$ approaches $R_{\text{SLB}}(D)$ as $D\to 0$.  
\begin{figure}[tb]
\centering
\resizebox{1.0\linewidth}{!}{
\begin{tikzpicture}
\begin{axis}[
 		xmode=log,
        ymin=1, ymax=5, 
        xmin=0.0001, xmax=1/3,
        tick label style={font=\small},
        ylabel style={rotate=-90},
        grid=major, 
        width=10cm, height=10cm,
        grid = major,
        grid style={gray!30},
        axis background/.style={fill=white},
        ylabel={$R$},
        xlabel={$D$},
        tick align=outside,
        legend entries={
				$R_\text{SLB}(D)$, 
				$R_\text{L}(D)$},
		legend style={legend pos=north east, font=\small}
     ]   
     \addplot+[darkgreen, no markers, dashed, line width=1pt] table[x expr=\thisrowno{0},y expr=\thisrowno{1}] {newshlb.dat};
     \addplot+[blue, no markers, line width=1pt] table[x expr=\thisrowno{0},y expr=(\thisrowno{1})] {shanlower2red.dat};
\end{axis}
\end{tikzpicture}
}
\caption{   
The Shannon lower bound $R_\text{SLB}(D)$ evaluated numerically in \cite[Section X.C]{kopirihl16} and the  lower bound  ${R_\text{L}}(D)$ in \eqref{eq:SLB1S1} for 
 $X$ distributed uniformly on the unit circle.\label{fig.bounds}}
\end{figure}
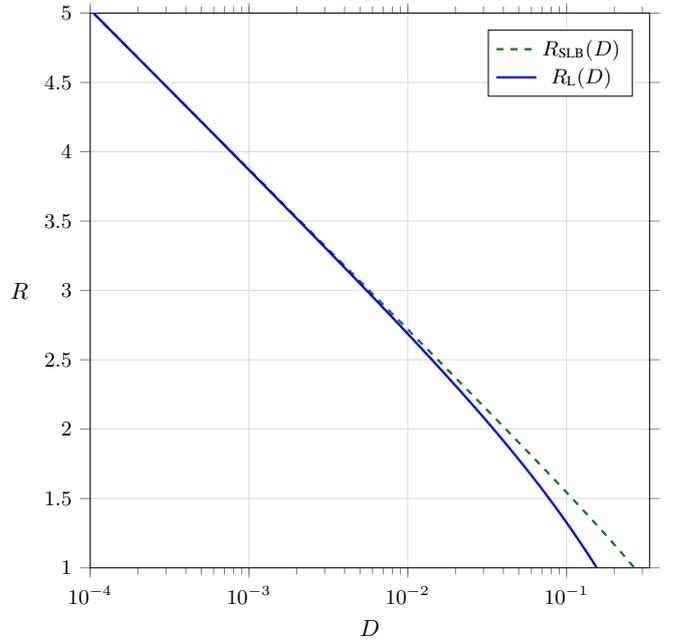
\end{exa}

To prepare the ground for  the second example, we need some preliminaries on contracting similarities; we follow the exposition in \cite{frheolro15}. 
A mapping $s\colon \reals^d\to\reals^d$ is called a contracting similarity if there exists a $\kappa \in (0,1)$, referred to as contraction parameter, such that 
\begin{align}
\lVert s(\vecu)-s(\vecv)\rVert_2= \kappa\lVert\vecu-\vecv\rVert_2\quad \text{for all}\ \vecu,\vecv\in\reals^d. \label{eq:contractions}
\end{align} 
For  $i\in\setI:=\{1,\dots,|\setI|\}$, consider contracting similarities $s_i\colon \reals^d\to\reals^d$  with corresponding contraction parameters $\kappa_i\in(0,1)$.   
By \cite[Theorem 9.1]{fa90}, there exists a unique  self-similar set 
\be 
\setK=\bigcup_{i\in\setI} s_i(\setK)\,\subseteq\reals^d. 
\ee
Let $\setI^\ast=\bigcup_{j\in\naturals}\setI^j$.  
For every $\alpha=(i_1,\dots,i_j)\in\setI^\ast$, we set $\bar\alpha=(i_1,\dots,i_{j-1})\in\setI^\ast\cup\{\omega\}$
with $\omega$ denoting the  empty sequence of length zero.  
We designate the identity mapping on $\reals^d$ by $s_\omega$, set $\kappa_\omega=1$,  
and define 
\begin{align}
s_\alpha&=s_{i_1}\circ s_{i_2}\circ\dots\circ s_{i_j}\\
\kappa_\alpha&=\kappa_{i_1}\kappa_{i_2}\dots \kappa_{i_j}\,
\end{align}   
for all $\alpha\in\setI^\ast$. 
It follows directly that  $s_\alpha$ is a contracting similarity with  contraction parameter $\kappa_\alpha$ for all $\alpha\in\setI^\ast$.  
Finally, for every $\delta>0$ and $x\in\setX$, let 
\begin{align}
\setJ_\delta&=\{\alpha\in\setI^\ast:\kappa_\alpha\leq \delta<\kappa_{\bar\alpha}\}\\
\setJ_\delta(x)&=\mleft\{\alpha\in\setJ_\delta:\setB_{\lVert\,\cdot\,\rVert_2}\big(x,\delta \big)\cap s_\alpha(\setK)\neq\emptyset\mright\}. \label{eq:setJd}
\end{align}
The following result will allow us to establish subregularity for  random variables distributed uniformly on  self-similar sets.  
  
\begin{lem}\cite[Theorem 2.1]{frheolro15}\label{lem:frheolro15}
For  $i\in\setI:=\{1,\dots,|\setI|\}$, consider contracting similarities $s_i\colon \reals^d\to\reals^d$  with  contraction parameters $\kappa_i\in(0,1)$. Let  
\be 
\setK=\bigcup_{i\in\setI} s_i(\setK) 
\ee
be the corresponding self-similar set and let $m$  be the similarity dimension given by the unique solution of 
\be
\sum_{i=1}^k \kappa_i^m=1. 
\ee
Then, 
\begin{align}
\colH^m\mleft(\setB_{\lVert\,\cdot\,\rVert_2}\big(x,\delta\big)\mright)
&\leq \colH^m(\setK)|\setJ_\delta(x)|\delta^m\label{eq:cantorsub2a}
\end{align}
for all $x\in\reals^d$ and $\delta\in (0,\infty)$.
If, in addition, the contracting similarities satisfy the weak separation property \cite[Definition on p. 3533]{ze96}  and $\setK$ 
is not contained in any   hyperplane of dimension $d-1$, then $0<\colH^m(\setK)<\infty$ 
and 
\begin{equation}
\colH^m\mleft(\setB_{\lVert\,\cdot\,\rVert_2}\big(x,\delta\big)\mright)
\leq c \delta^m\quad \text{for all $x\in\reals^d$ and $\delta\in (0,\infty)$}  \label{eq:cantorsub2}
\end{equation}
with  $c>1$ and independent of $x$ and $\delta$. 
\end{lem}

We are now ready to present our second example, namely, a random variable distributed  uniformly on the middle third Cantor set. 
\begin{exa}(Uniform distribution on the middle third Cantor set)\label{ex:CA}
Let $\setX=\setY=\reals$ be equipped with the Borel $\sigma$-algebra and the distortion function $\rho(x,y)=\lVert x-y\rVert_2^2$. 
Consider the middle third Cantor set $\setC\subseteq [0,1]$, i.e., the self-similar  set corresponding to  $\setI=\{1,2\}$, $\kappa_1=\kappa_2=1/3$, $s_1(x)=x/3$, $s_2(x)=x/3+2/3$, and $m=\log 2/\log 3$. 
Since   $0<\mathscr{H}^{\log 2/\log 3}(\setC)<\infty$  \cite[Example 4.5]{fa90}, we can take 
 $X$ distributed uniformly on $\setC$, i.e., $\mu_X=\colH^m|_\setC/\colH^m(\setC)$.  
Next, we use \eqref{eq:cantorsub2a} in Lemma \ref{lem:frheolro15} to obtain a subregularity condition for $\mu=\mu_X$. To this end, it is first shown that $|\setJ_{\delta}(x)|\leq 3$ for all $\delta \in (0,1)$ and $x\in \reals$.
Note that  
$\kappa_\alpha=3^{-j}$ for  all $\alpha=(i_1,\dots,i_j)$ and  $j\in\naturals_0$.  Thus,    
\begin{align}
\setJ_{\delta}
&=\{\alpha\in\setI^\ast:\kappa_\alpha\leq \delta <\kappa_{\bar\alpha}\}\\
&=\{\alpha:|\alpha|=j\}\quad\text{for all $\delta\in\big[3^{-j},3^{-j+1}\big)$ and $j\in\naturals$},  
\end{align}
which implies $|\setJ_{\delta}(x)|\leq 3$ for all $\delta\in (0,1)$ and $x\in\reals$ (cf. Figure \ref{fig:cantor}). 
\begin{figure}[tb]
\resizebox{0.77\linewidth}{!}{
\begin{tikzpicture}[scale=2]
  \foreach \order in {0,...,4}
    \draw[line width=0.5mm, yshift=-\order*10pt]  l-system[l-system={cantor set, axiom=F, order=\order, step=100pt/(3^\order)}];
\put (220,-22){$|\alpha|=1$};
\put (220,-42){$|\alpha|=2$};  
\put (220,-62){$|\alpha|=3$};   
\put (220,-82){$|\alpha|=4$};
\put (10,0){\draw[{Arc Barb[]}-{Arc Barb[]}, ultra thick, red, dashed] (0,-20pt) -- (66pt,-20pt);};
\put (0,0){
\put (4,-90){\draw [dotted, line width=0.3mm] (0pt,0pt) -- (0pt,-7pt);};
\put (18,-90){\draw [dotted, line width=0.3mm] (0pt,0pt) -- (0pt,-7pt);};
};
\put (45,0){
\put (4,-90){\draw [dotted, line width=0.3mm] (0pt,0pt) -- (0pt,-7pt);};
\put (18,-90){\draw [dotted, line width=0.3mm] (0pt,0pt) -- (0pt,-7pt);};
};
\put (134,0){
\put (4,-90){\draw [dotted, line width=0.3mm] (0pt,0pt) -- (0pt,-7pt);};
\put (18,-90){\draw [dotted, line width=0.3mm] (0pt,0pt) -- (0pt,-7pt);};
};
\put (178,0){
\put (4,-90){\draw [dotted, line width=0.3mm] (0pt,0pt) -- (0pt,-7pt);};
\put (18,-90){\draw [dotted, line width=0.3mm] (0pt,0pt) -- (0pt,-7pt);};
};
\end{tikzpicture}
}
\vspace*{10truemm}
\caption{Sets $s_\alpha([0,1])$ with $|\alpha|=j$ have length $3^{-j}$. At most three different  sets $s_\alpha([0,1])$ with $|\alpha|=j$ intersect with an open interval of  length $2(3^{-j+1})$. 
\label{fig:cantor}} 
\end{figure}
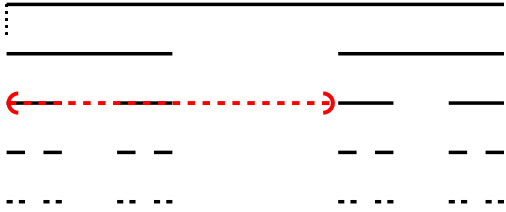
Therefore, 
\eqref{eq:cantorsub2a} together with $m=\log 2/\log 3$  yields the subregularity condition 
\begin{align}
\mu_X\mleft(\setB_{\lVert\,\cdot\,\rVert_2}\big(x,\delta\big)\mright)&\leq 3 \delta^\frac{\log 2}{\log 3}\quad \text{for all  $x\in\reals$ and $\delta\in (0,\infty)$}.  \label{eq:Cantorsub}
\end{align}
With  \eqref{eq:Cantorsub}  
the lower bound $R_\text{L}(D)$ in 
 \eqref{eq:SLB1} for $\mu=\mu_X$,  $m=\log2/\log3$, $k=2$, $\delta_0=\infty$, and  $c=3$ 
and hence $c\geq \mu_X(\setX)/\delta_0=0$ is given by  
\begin{equation}\label{eq:SLB1aC}
R_{\text{L}}(D) = 
\sigma \log \mleft(\frac{\sigma}{D} \mright)-\sigma
-\log
\mleft(3
\Gamma\mleft(\sigma+1\mright)
\mright)\quad\text{for all $D>0$}, 
\end{equation}
where $\sigma:=\log2/\log9$. 
\end{exa}
\bibliographystyle{IEEEtran}
\bibliography{references}

\begin{thebibliography}{10}
\providecommand{\url}[1]{#1}
\csname url@samestyle\endcsname
\providecommand{\newblock}{\relax}
\providecommand{\bibinfo}[2]{#2}
\providecommand{\BIBentrySTDinterwordspacing}{\spaceskip=0pt\relax}
\providecommand{\BIBentryALTinterwordstretchfactor}{4}
\providecommand{\BIBentryALTinterwordspacing}{\spaceskip=\fontdimen2\font plus
\BIBentryALTinterwordstretchfactor\fontdimen3\font minus
  \fontdimen4\font\relax}
\providecommand{\BIBforeignlanguage}[2]{{%
\expandafter\ifx\csname l@#1\endcsname\relax
\typeout{** WARNING: IEEEtran.bst: No hyphenation pattern has been}%
\typeout{** loaded for the language `#1'. Using the pattern for}%
\typeout{** the default language instead.}%
\else
\language=\csname l@#1\endcsname
\fi
#2}}
\providecommand{\BIBdecl}{\relax}
\BIBdecl

\bibitem{bawa09}
R.~G. Baraniuk and M.~B. Wakin, ``Random projections of smooth manifolds,''
  \emph{Found. Comput. Math.}, vol.~9, no.~1, pp. 51--77, Feb 2009.

\bibitem{care09}
E.~J. Cand\`{e}s and B.~Recht, ``Exact matrix completion via convex
  optimization,'' \emph{Found. Comput. Math.}, vol.~9, no.~6, pp. 717--772,
  2009.

\bibitem{capl11}
E.~J. Cand\`{e}s and Y.~Plan, ``Tight oracle inequalities for low-rank matrix
  recovery from a minimal number of noisy random measurements,'' \emph{{IEEE}
  {T}rans. {I}nf. {T}heory}, vol.~4, no.~57, pp. 2342--2359, Apr. 2011.

\bibitem{albdekori18}
G.~Alberti, H.~B\"olcskei, C.~De~Lellis, G.~Koliander, and E.~Riegler,
  ``Lossless analog compression,'' \emph{submitted to {IEEE} {T}rans. {I}nf.
  {T}heory, arXiv:1803.06887}, 2018.

\bibitem{ristbo15}
E.~Riegler, D.~Stotz, and H.~B\"olcskei, ``Information-theoretic limits of
  matrix completion,'' in \emph{{P}roc. {IEEE} {ISIT}}, Jun. 2015, pp.
  1836--1840.

\bibitem{lz12}
A.~J. Lzenman, ``Introduction to manifold learning,'' \emph{WIREs Comput.
  Stat.}, vol.~4, pp. 439--446, 2012.

\bibitem{lufahe98}
H.~Lu, Y.~Fainman, and R.~Hecht-Nielsen, ``Image manifolds,'' in \emph{Proc.
  SPIE}, San Jose, CA, Apr 1998, pp. 3307--3307.

\bibitem{soze98}
N.~Sochen and Y.~Y. Zeevi, ``Representation of colored images by manifolds
  embedded in higher dimensional non-{E}uclidean space,'' in \emph{{P}roc.
  {IEEE} {ICIP}}, Chicago, IL, Oct 1998, pp. 166--170.

\bibitem{hidare97}
G.~E. Hinton, P.~Dayan, and M.~Revow, ``Modeling the manifolds of images of
  handwritten digits,'' \emph{IEEE Trans. Neural Netw.}, vol.~8, no.~1, pp.
  65--74, Jan 1997.

\bibitem{letawi94}
W.~E. Leland, M.~S. Taqqu, W.~Willinger, and D.~V. Wilson, ``On the
  self-similar nature of {E}thernet traffic (extended version),''
  \emph{IEEE/ACM Trans. Netw.}, vol.~2, no.~1, pp. 1--15, Feb 1994.

\bibitem{sh59}
C.~E. Shannon, ``Coding theorems for a discrete source with a fidelity
  criterion,'' \emph{IRE Nat. Conv. Rec. Pt. 4}, vol.~7, pp. 142--163, 1959.

\bibitem{be71}
T.~Berger, \emph{{R}ate {D}istortion {T}heory: A Mathematical Basis for Data
  Compression}.\hskip 1em plus 0.5em minus 0.4em\relax Englewood Cliffs, NJ:
  Prentice-Hall, 1971.

\bibitem{gr90}
R.~M. Gray, \emph{Source Coding Theory}.\hskip 1em plus 0.5em minus 0.4em\relax
  Boston, MA: Kluwer, 1990.

\bibitem{grne98}
R.~M. Gray and D.~L. Neuhoff, ``Quantization,'' \emph{{IEEE} {T}rans. {I}nf.
  {T}heory}, vol.~44, no.~6, pp. 2325--2383, Oct. 1998.

\bibitem{gr11}
R.~M. Gray, \emph{Entropy and Information Theory}, 2nd~ed.\hskip 1em plus 0.5em
  minus 0.4em\relax New York, NY: Springer, 2011.

\bibitem{kade94}
T.~Kawabata and A.~Dembo, ``The rate-distortion dimension of sets and
  measures,'' \emph{{IEEE} {T}rans. {I}nf. {T}heory}, vol.~40, no.~5, pp.
  1564--1572, Sep. 1994.

\bibitem{ro88}
H.~Rosenthal, ``On the epsilon entropy of mixed random variables,''
  \emph{{IEEE} {T}rans. {I}nf. {T}heory}, vol.~34, no.~5, pp. 1110--1114, Sep.
  1988.

\bibitem{yatagr80}
Y.~Yamada, S.~Tazaki, and R.~M. Gray, ``Asymptotic performance of block
  quantizers with difference distortion measure,'' \emph{{IEEE} {T}rans. {I}nf.
  {T}heory}, vol.~26, no.~1, pp. 6--14, Jan. 1980.

\bibitem{liza94}
T.~Linder and R.~Zamir, ``On the asymptotic tightness of the {S}hannon lower
  bound,'' \emph{{IEEE} {T}rans. {I}nf. {T}heory}, vol.~40, no.~6, pp.
  2026--2031, Nov. 1994.

\bibitem{ko16}
T.~Koch, ``The {S}hannon lower bound is asymptotically tight,'' \emph{{IEEE}
  {T}rans. {I}nf. {T}heory}, vol.~62, no.~11, pp. 6155--6161, Nov. 2016.

\bibitem{cs74}
I.~Csisz\'ar, ``On an extremum problem of information theory,'' \emph{Stud.
  Sci. Math. Hung.}, no.~9, pp. 57--71, 1974.

\bibitem{kopirihl16}
G.~Koliander, G.~Pichler, E.~Riegler, and F.~Hlawatsch, ``Entropy and source
  coding for integer-dimensional singular random variables,'' \emph{{IEEE}
  {T}rans. {I}nf. {T}heory}, vol.~62, no.~11, pp. 6124--6154, 2016.

\bibitem{ko17}
V.~Kostina, ``Data compression with low distortion and finite blocklength,''
  \emph{{IEEE} {T}rans. {I}nf. {T}heory}, vol.~63, no.~7, pp. 4268--4285, Jul.
  2017.

\bibitem{grlu00}
S.~Graf and H.~Luschgy, \emph{Foundations of Quantization for Probability
  Distributions}.\hskip 1em plus 0.5em minus 0.4em\relax Berlin, Germany:
  Springer, 2000.

\bibitem{frheolro15}
J.~M. Fraser, A.~M. Henderson, E.~J. Olson, and J.~C. Robinson, ``On the
  dimension of self-similar sets with overlaps,'' \emph{Adv. Math.}, vol. 273,
  pp. 188--214, 2015.

\bibitem{amfupa00}
L.~Ambrosio, N.~Fusco, and D.~Pallara, \emph{Functions of {B}ounded {V}ariation
  and {F}ree {D}iscontinuity {P}roblems}.\hskip 1em plus 0.5em minus
  0.4em\relax New York, NY: Oxford Univ. Press, 2000.

\bibitem{fa90}
K.~Falconer, \emph{Fractal Geometry}, 1st~ed.\hskip 1em plus 0.5em minus
  0.4em\relax New York, NY: Wiley, 1990.

\bibitem{ze96}
M.~P.~W. Zerner, ``Weak separation properties for self-similar sets,''
  \emph{Proc. Am. Math. Soc.}, vol. 124, no.~11, pp. 3529--3539, 1996.

\end{thebibliography}
\end{document}